\documentclass[aps,pre,reprint,groupedaddress,amssymb]{revtex4-2}

\makeatletter
\def\set@curr@file#1{%
\begingroup
\escapechar\m@ne
\xdef\@curr@file{\expandafter\string\csname #1\endcsname}%
\endgroup
}
\def\quote@name#1{"\quote@@name#1\@gobble""}
\def\quote@@name#1"{#1\quote@@name}
\def\unquote@name#1{\quote@@name#1\@gobble"}
\makeatother
\usepackage{graphicx}% Include figure files
\usepackage{dcolumn}% Align table columns on decimal point
\usepackage{bm}% bold math
\usepackage{amsmath}%Align equations
\usepackage{xcolor}
\usepackage{hyperref}
\usepackage{nicefrac}
\usepackage{caption}
\usepackage{subcaption}
\usepackage{epstopdf}
\usepackage{tikz}
\usepackage{caption}
\usepackage{nicefrac}
\usepackage{lipsum}
\usepackage{cancel}
\usepackage[toc,page]{appendix}
\usepackage[inline]{enumitem}   
\usepackage [autostyle, english = american]{csquotes}
\MakeOuterQuote{"}
\usepackage[margin=10pt,font=small,labelfont=bf,justification=centerlast,format=plain]{caption}
\begin{document}

%\title{
%	No Name
	%Naive mean-field approximation and simultaneous updating %
%}

%\title{Modified Heider Balance on Sparse Random Networks}
%\author{R. Masoumi\textsuperscript{1}}
%\author{F. Oloomi\textsuperscript{1}}
%\author{S. Sajjadi\textsuperscript{2,3}}
%\author{A.H. Shirazi\textsuperscript{1}}
%\email{al\_hosseiny@sub.ac.ir}
%\author{G. R. Jafari\textsuperscript{1,4}}
%\email{g\_jafari@sbu.ac.ir}	

%\affiliation{\textsuperscript{1} Department of Physics, Shahid Beheshti University, Evin, Tehran 19839, Iran}
%\affiliation{\textsuperscript{2} Complexity Science Hub Vienna, Vienna, Austria}
%\affiliation{\textsuperscript{3} Central European University, Vienna, Austria}
%\affiliation{\textsuperscript{4} Institute of Information Technology and Data Science, Irkutsk National Research Technical University, 83, Lermontova St., 664074 Irkutsk, Russia}

\title{Modified Heider Balance on Sparse Random Networks
}
\author{R. Masoumi$^{1}$}
%\thanks{Both authors contributed equally to this work.}
\author{F. Oloomi$^{1}$}
%\thanks{Both authors contributed equally to this work.}
\author{S. Sajjadi$^{2,3}$}
%\email[]{fakhteh.ghanbarnejad@gmail.com}
\author{A.H. Shirazi$^{1}$}
\author{G.R. Jafari$^{1,4}$}
\email{g\_jafari@sbu.ac.ir}	
\affiliation{
	$^1$Department of Physics, Shahid Beheshti University, Evin, Tehran 19839, Iran\\
	$^2$Complexity Science Hub Vienna, Vienna, Austria\\
	$^3$Central European University, Vienna, Austria\\
	$^4$Institute of Information Technology and Data Science, Irkutsk National Research Technical University, 83, Lermontova St., 664074 Irkutsk, Russia
}

\date{\today}
\keywords{Keyword1, Keyword2, Keyword3}

\begin{abstract}
The lack of signed random networks in standard balance studies has prompted us to extend the Hamiltonian of the standard balance model.
Random networks with tunable parameters are suitable for better understanding the behavior of standard balance as an underlying dynamics. Moreover, the standard balance model in its original form does not allow preserving tensed triads in the network. Therefore, the thermal behavior of the balance model has been investigated on a fully connected signed network recently. It has been shown that the model undergoes an abrupt phase transition with temperature. Considering these two issues together, we examine the thermal behavior of the structural balance model defined on Erd\H{o}s-R\'enyi random networks. We provide a Mean-Field solution for the model. We observe a first-order phase transition with temperature, for both the sparse and densely connected networks. We detect two transition temperatures, $T_{cold}$ and $T_{hot}$, characterizing a hysteresis loop. We find that with increasing the network sparsity, both $T_{cold}$ and $T_{hot}$ decrease. But the slope of decreasing $T_{hot}$ with sparsity is larger than the slope of decreasing $T_{cold}$. Hence, the hysteresis region gets narrower, until, in a certain sparsity, it disappears. We provide a phase diagram in the temperature-tie density plane to observe the meta-stable/coexistence region behavior more accurately. Then we justify our Mean-Field results with a series of Monte-Carlo simulations.

\end{abstract}

\maketitle

\section*{Introduction}

Network models are a powerful method for describing complex phenomena. Networks represent systems as a set of nodes and ties between them, where nodes denote entities and ties represent a type of association between a pair of entities. This association can be friendship in social relations, a transaction in financial networks, or the possibility of infection in an epidemiological setting.
%This association can be friendship in social relations, a transaction in financial networks, or the possibility of infection in an epidemiological setting.
While simple networks can represent the cooperation, alliance, friendship, communication, trust, and correlation between nodes, they lack the capability of incorporating the notion of rivalry, conflict, enmity, distrust, or negative correlation. Signed networks bring this possibility into a network model by introducing signed ties, with positive ties representing the former and negative ties representing the latter types of relations \cite{szell,singh,Altafini2011}.
%\cite{ecol2017,Altafini2011,Davis1967, Singh2016, Doreian2015, Bramson2017, Galam1996, estrada1, Thurner2020}.
Therefore, the signed network method has found many applications in various disciplines ranging from sociology \cite{Altafini2012,kulakowski2019,Thurner2020,Montakhab}, epidemiology \cite{Epidemic2017}, international relations \cite{hart,galam,bramson,estrada1,Doreian2015}, politics \cite{Aref2020}, and ecology \cite{ecol2017}.

Signed networks have been employed by the structural balance theory \cite{Heider1946, Cartwright1956}, to study the equilibrium states in networks with both negative and positive associations.
According to this theory, the state of balance for a network is defined based on the status of its triads, motifs consisting of three nodes with three ties connecting them.
A triad is defined as balanced or non-tensed if an even number of its ties are negative ($[+,+,+]$, $[-,-,+]$). Otherwise, it is considered an unbalanced or tensed triad ($[-,+,+]$, $[-,-,-]$) \cite{Heider1946}.
Intuitively, thinking of signs as friendship/enmity, it can be shown that balance holds when the following statements hold for all nodes in a triad:
\begin{enumerate*}
	\item \textit{The friend of my friend is my friend.}
	\item \textit{The friend (enemy) of my enemy (friend) is my enemy.}
	\item \textit{The enemy of my enemy is my friend.}
\end{enumerate*}

In the simplest versions of the model, tie signs get updated until the system reaches a fully balanced state. In this state, the network comprises two communities, where all intra-community and inter-community ties are respectively positive and negative.
The classic works done on the dynamics of structural balance model are \cite{Antal2005, kulakowski2005, Marvel2010}.
However, only rarely do real-world social networks arrive at a fully balanced state. While balanced triads are more prevalent, unbalanced triads still exist.
To reflect this issue, a relaxed version of the theory has been devised by introducing a source of randomness via the adoption of the concept of the \textit{social temperature} \cite{Amirhossein}.
By doing so, the structural balance model has been mapped to a Boltzmann-Gibbs statistical model where each state is assigned with a probability $\frac{e^{\nicefrac{-{E}}{{T}}} }{C}$. Here $E$ denotes the energy of the system in this state, $T$ represents the social temperature and $C$ is the normalization factor. In this picture, the energy, ${E}_{\Delta}$, assigned to balanced and unbalanced triads is respectively $-1$ and $+1$ so that states with a higher number of balanced triads are assigned with a high probability.
$T$ represents the level of tension tolerance in the system.
With $T \rightarrow 0$ one would achieve the strict model in which only balanced triads are allowed and $T \rightarrow \infty$ would lead to a system, neutral toward triadic un/balance.

%This model has been solved via a .
%As another consequence arising from the intrinsic feature of the discontinuous phase transition, there is a bi-stable region in which both random and balance phases coexist.
%Therefore, if we quench the system to the temperature below $T_{c}$, the system stays in its random phase. From an intuitive point of view, it is somewhat surprising to see the systems in a random phase below the tension tolerance threshold. However, it should be noted that the fixed point representing the random phase within the coexistence region has a very narrow basin of attraction, thus this random state is extremely vulnerable.

Most analytical studies of the structural balance model have been conducted on complete graphs, neglecting the underlying network structures \cite{Amirhossein, Amir, Raha, Farideh, Mahsa, Hassanibesheli}.
%Using a Mean-Field approximation for complete graphs, it has been observed that there exists a temperature threshold $T_{c}$ where the system undergoes a discontinuous phase transition from a structural balance state to a random state \cite{Amirhossein}. \Sina{Structural balance state? Or balanced state? And why do we need to mention this result here in the introduction?}
%In other words, there is no chance for the system to reach the state of balance with $T>T_{c}$
While empirical signed networks have also been employed \cite{Altafini2011, Belaza2017, Belaza2019}, random networks have not attracted much attention in structural balance models. Random networks with controllable parameters have helped in understanding the effects of network characteristics on the dynamics of different phenomena such as spreading and percolation \cite{Vespignani1, Vespignani2, Newman}. Hence, a similar methodology can be helpful in the study of the network aspects of the structural balance. Recently, the thermal behavior of structural balance has been investigated on diluted and enhanced triangular lattices by performing a series of simulations \cite{KrzysztofMalarz}.

Some random network models have been introduced to describe the real-world networks better. See for example Erd\H{o}s-R\'enyi random graphs, Watts Strogatz Small-world networks and growing random graphs like preferential attachment network of Barabasi Albert,\cite{barabasi}.

%In the case of diluted triangular lattice, where the mean of node degree is so close to node degree in triangular lattice reaching the state of balance is impossible for all temperatures.}
%The first one is a diluted triangular lattice that is created by removing some ties from a triangular lattice or an enhanced triangular lattice that is created by adding more ties to a triangular lattice. The second one is the classical random graph.
%They have discusses all possible range of triangle densities in both cases, from a fully connected network to a very sparse networks with some few disconnected triangles.

%\textcolor{blue}{Bringing these two ideas (tension tolerance and random network) together, Now we are able to manipulate our random network's parameters to capture its dynamical behavior in the presence of different levels of tension tolerance. Among random networks, Erd\H{o}s-R\'enyi random graphs are the most simple type of random graphs. Although they have a Poissonian degree distribution in the limit of large size, however, they are classified in the category of spars networks. Is this pararaph necessary? or can we say it in a better way? }

In this study, we first introduce a Hamiltonian for the structural balance model defined on a class of random graphs called Erd\H{o}s-R\'enyi graphs. Then, we investigate the stationary states of our model in the presence of social temperature.
We present a Mean-Field solution for our model under a canonical ensemble. We observe that by varying the temperature, the system undergoes a discontinuous phase transition.
We detect two transition temperatures which we name $T_{cold}$ and $T_{hot}$.
% As their names imply the cold temperature $T_{cold}$ is where, before that, the system is completely in a balanced phase. On the other hand, the hot temperature is where, after that, the system is completely in a random phase.
For $T < T_{cold}$ and $T > T_{hot}$ the system would respectively settle in a completely balanced and a completely random phase. For $T_{cold} < T < T_{hot}$ the system undergoes a bi-stability phase experiencing both random and balanced phases.
We calculate $T_{cold}$ analytically and show that the coexistence region gets narrower as the connection probability decreases. Finally, we perform a series of Monte-Carlo simulations to support our Mean-Field solutions.

\section*{Model}
In this section, we present a Mean-Field solution for the structural balance model defined on Erd\H{o}s-R\'enyi networks.
An Erd\H{o}s-R\'enyi network, is a static random network with a fixed number of nodes, in which each tie exists with an identical independent probability $p$ \cite{Erdos}.

Inspired by the structural balance Hamiltonian, we define the Hamiltonian of our model on Erd\H{o}s-R\'enyi networks in Eq. \ref{eq:1}.

\begin{equation}\label{eq:1}
\textbf{H}=-\frac{1}{N}\sum_{i<j<k} s_{ij} s_{jk} s_{ki} e_{ij} e_{jk} e_{ki},
\end{equation} 

In Eq. \ref{eq:1}, $N$ is the number of nodes, and $s_{ij} \in \{+1, -1\}$ indicates the relationship between nodes $i$ and $j$. Network topology is encoded in the adjacency matrix $e$, where $e_{ij} = 1$ if nodes $i$ and $j$ are connected, and $e_{ij} = 0$ otherwise.
To evaluate the model, we need to investigate its observable/macroscopic quantities.
The macroscopic quantities in our model are ensemble averages of ties, two-stars, and energy, respectively denoted by $\langle s_{ij}\rangle$, $\langle s_{ik} s_{kj}\rangle$ and -$\langle s_{ij} s_{jk} s_{ki}\rangle$.
The role of the mean of two-stars on the dynamic of thermal balance has been studied in \cite{Amirhossein}. This quantity measures the closeness of a network to the balanced state.
%\Sina{I edited this part, but I think you have neglected my comment here. First, Why do you only explain one of the three quantities. Second, Then isn't this sentence describing the energy? Low energy -> closer to balanced state?}
%\textcolor{blue}{Let's take a simple example to make the concept of mean of two-stars a little more concrete. When a fully connected signed network is balanced its favored configurations are utopia or bipolar. So $\langle s_{ik} s_{kj}\rangle$ or better to say the correlation between two ties that have a node in common reduces to the number of triangles established on each tie which has its maximum value in these cases. On the other hand, when ties' signs are distributed randomly there is no correlation between ties in a two-star, so $\langle s_{ik} s_{kj}\rangle =0 $. Hence this quantity is a measure of up to what extent a network is close to balanced configurations.}

To calculate these quantities, we require the probability distribution of the possible states of the system.
For this purpose, we define a partition function for our Hamiltonian \eqref{eq:1}, in a canonical ensemble of the $s$ variable. The partition function is written as in Eq. \ref{eq:2}.
\begin{equation}\label{eq:2}
\textbf{Z}=\sum_{\{\textbf{s}\}} e^{-\beta \textbf{H}}.
\end{equation} 
The $\{\textbf{s}\}$ subscript indicates taking the summation over the ensemble of all possible signed ties in a static Erd\H{o}s-R\'enyi network. 
%This method is identical to the \textit{quenched disorder} in spin glass models, where some random variables are quenched and do not evolve in time but their whole range will be spanned in the integration \cite{Nishimori}.
In this study, we assume that the network structure is the quenched/frozen variable. Therefore, to calculate the free energy, we trace
over all possible graph configurations.
The free energy obtained using this approach is called the \textit{configurational free energy}.
%\Sina{Citation needed.}

To calculate the ensemble average of a tie sign, $\langle s_{ij} \rangle$, we split the Hamiltonian in two parts (Eq. \ref{eq:3}), $ \textbf{H}_{ij} $ consisting of the terms which the tie $\{i,j\}$ contributes to (Eq. \ref{eq:4}), and $ {\textbf{H}}^{\prime}$ comprising the rest of the terms.

%To obtain the average of ties, $\langle s_{ij} \rangle$, in our Mean-Field approximation, we first isolate existing tie $e_{ij}=1$ and subsequently $s_{ij}$ which is pinned to. Then we consider the remaining existing ties as an effective field acting on $s_{ij}$. For this purpose, we rewrite the Hamiltonian as follows:

\begin{eqnarray}\label{eq:3}
\textbf{H}=\textbf{H}_{ij}+ {\textbf{H}}^{\prime}
\end{eqnarray} 
%where, $\textbf{H}_{ij}$ includes all terms of the Hamiltonian which contains tie $\{i,j\}$ and $\textbf{H}^{\prime}$ includes the remaining terms of the Hamiltonian which do not contain tie $\{i,j\}$. So $\textbf{H}_{ij}$ is as below:
\begin{eqnarray}  \label{eq:4}
\textbf{H}_{ij} = -\frac{1}{N} s_{ij} \sum_{k\neq i,j} s_{jk}s_{ki} e_{jk} e_{ki} -h_{ij}s_{ij},
\end{eqnarray}
Where $h_{ij}$ is considered as a external field on $s_{ij}$.
Therefore, the partition function can be written as follows:
\begin{equation} \label{eq:45}
\begin{aligned}
\textbf{Z}&=\sum_{\{\textbf{s}\}} e^{-\beta \textbf{H}}= \sum_{\{\textbf{s}\}} e^{-\beta ({\textbf{H}}_{ij}+{\textbf{H}}^{\prime})}\\
&={\textbf{Z}}^{\prime} \sum_{\{\textbf{s}\neq s_{ij}\}} \frac{e^{-\beta {{\textbf{H}}}^{\prime}}}{{\textbf{Z}}^{\prime}} \sum_{s_{ij}= \{ \pm 1\}} e^{-\beta {\textbf{H}}_{ij}}
={\textbf{Z}}^{\prime} {\bigg\langle  \sum_{s_{ij}= \{ \pm 1\}} e^{-\beta {\textbf{H}}_{ij}}\bigg \rangle}_{Z^{\prime}}\\
&={\textbf{Z}}^{\prime} {\bigg \langle 2\cosh \bigg(  \frac{\beta}{N} \sum_{k\neq i,j} s_{jk} s_{ki} e_{jk}e_{ki} + \beta h_{ij} \bigg)  \bigg \rangle}_{Z^{\prime}}\\
&=2 {\textbf{Z}}^{\prime} \cosh \bigg(\frac{\beta}{N} {\bigg \langle \sum_{k \neq i,j} s_{jk} s_{ki} e_{jk}e_{ki} \bigg\rangle}_{Z^{\prime}} + \beta h_{ij} \bigg)
\end{aligned}
\end{equation}
%As we discussed the partition function is quenched respect to $\enquote e$ variable, Hence just $\enquote s$ variable is allowed to fluctuate. So, the expression $\bigg\langle \sum_{k} e_{jk} e_{ki}s_{jk}s_{ki} \bigg\rangle$ is assumed to be approximated by the number of two-stars named $m$ that are established on tie $\{i,j \}$, multiplied by the mean of two-stars $\langle s_{jk}s_{ki} \rangle$. For making this approximation we have used the homogeneity assumption for the Erd\H{o}s-R\'enyi random graph. So we have:

As a result of the Mean-Field approximation, the term $\bigg\langle \sum_{k \ne i, j}s_{jk}s_{ki} e_{jk} e_{ki} \bigg\rangle$ can be approximated by $m$, the number of triangles that two-stars make with tie  $\{i,j \}$, multiplied by the ensemble average of the two-stars $\langle s_{jk}s_{ki} \rangle$. So we have:
%\Sina{Then its just the number of triangles consisting of tie $\{i,j \}$, why would we need to bring up the notion of two-stars in the definition of the triangles?}

%denoting the energy of the triangles consisting of tie $\{i,j \}$, can be approximated by $m$, the number of these triangles, multiplied by the ensemble average of the two-stars $\langle s_{jk}s_{ki} \rangle$. So we have:

\begin{eqnarray}\label{eq:5}
\textbf{Z}= \textbf{Z}_{ij}(m)\textbf{Z}^{\prime}= 2\cosh \bigg( \frac{\beta}{N} m {\bigg\langle ss \bigg \rangle}_{Z'} +\beta h_{ij} \bigg) {\textbf{Z}}^{\prime}
\end{eqnarray}
%In Eq.\eqref{eq:5} we have omitted index from the  $\langle s_{jk} s_{ki} \rangle $ quantity and it is logical because for a quenched configuration the mean of two-stars does not depend on index.
Where, we have decomposed the partition function into two parts $\textbf{Z}_{ij}$ and $\textbf{Z}^{\prime}$, respectively the parts which tie $\{i,j\}$ does and does not contribute to.
%For a specific graph configuration with existing tie $\{i,j\}$ $\textbf{Z}_{ij}=\textbf{Z}_{ij}(m)$ depends on $m$, where $m$ is the number of two-stars that are established on tie $\{i,j\}$.
%As in  Erd\H{o}s-R\'enyi random graph the tie between any two nodes are established randomly, so there is no correlation between different parts of this kind of graph.\\
%As a consequence, in configuration integral of free energy we can integrate over the part that includes tie $\{i,j\}$. On the other hand the remaining parts of the graph are considered as a whole which do not play a role in our calculations. \\
From statistical mechanics we have \cite{Amirhossein}: 
\begin{eqnarray}\label{eq:55}
\langle s_{ij} \rangle &=-\frac{\partial \big[\textbf{F}\big]_{c}}{\partial h_{ij}} {\Bigg|}_{h_{ij}=0}
\end{eqnarray}
So we have to calculate the \textit {configurational free energy} $[\textbf{F}]_{c}$, Where ${[\hspace*{2mm}]}_{c}$ indicates the integral over all possible random graph configurations.
As we can split the partition function in the form of $\textbf{Z}= \textbf{Z}_{ij}\textbf{Z}^{\prime}$, we can write the \textit{configurational free energy} as a sum of two terms, in which the first term represents the integration of the partition function $\textbf{Z}_{ij}$ over the terms including tie $\{i,j\}$ and the second term represents the integration of the partition function $\textbf{Z}^{\prime}$ over the rest. So we have:
\begin{equation}\label{eq:6}
\begin{aligned}
{[\textbf{F}]}_{c}&=-\frac{1}{\beta} \bigg([\textbf{F}^{\prime}]_{c^{\prime}} +  \sum_{c_{ij}} \ln \textbf{Z}_{ij}\bigg)\\
&=-\frac{1}{\beta} \bigg( [\textbf{F}^{\prime}]_{c^{\prime}} + \sum_{m}^{N-2} P(m)\ln \textbf{Z}_{ij}(m)\bigg)\\
\end{aligned}
\end{equation}
The details of the calculation can be found in Appendix. \ref{appendix:b}. The first term of Eq.\eqref{eq:6} does not play a role in our calculations, so we can consider it a constant.

As Eq. \eqref{eq:5} indicates, ${\textbf{Z}_{ij}}$ is a function of $m$, the number of triangles including the tie $\{i,j\}$, so the second term can be approximated by a summation over values of $m$. Defining $P(m)$ the probability of generation of $m$ triangles including the tie $\{i,j\}$, we have:

%So integral of ${\textbf{Z}_{ij}}$ over part $\{c_{ij}\}$ could be approximated by summing over all possible states of constructing two-stars on tie $\{i,j\}$. $P(m)$ is the probability of establishing $m$ two-stars on a tie like $\{i,j\}$ which is:

\begin{equation}\label{eq:7}
P(m)= \binom {N-2}{m} {(p^2)}^{m} \left( \left(1-p\right)^2+2p\left(1-p\right) \right)^{N-2-m} 
\end{equation} 
Therefore, ${[\textbf{F}]}_{c}$ will be:
\begin{equation}\label{eq:8}
\begin{aligned}
{[\textbf{F}]}_{c}
% &= -\frac{A[\textbf{F}^{\prime}]_{c^{\prime}}}{\beta} -\frac{1}{\beta}\sum_{m}^{N-2} P(m) \ln\bigg(\cosh \big( \frac{\beta }{N} m  {\big\langle ss \big \rangle}_{\textbf{Z}^{\prime}} +\beta h_{ij} \big)\bigg)\\
&=-\frac{[\textbf{F}^{\prime}]_{c^{\prime}}}{\beta}-\frac{1}{\beta} \ln\bigg(\cosh \big( \frac{\beta }{N} (N-2)p^{2}  {\big\langle ss \big \rangle}_{\textbf{Z}^{\prime}} +\beta h_{ij} \big)\bigg)
\end{aligned}
\end{equation}
%Now we substitute Eq.\eqref{eq:5} in Eq.\eqref{eq:6} to obtain the configuration free energy
\begin{figure*}[t!]
	\begin{subfigure}{0.48\linewidth}
		\includegraphics[width=\linewidth, height=0.6\linewidth]{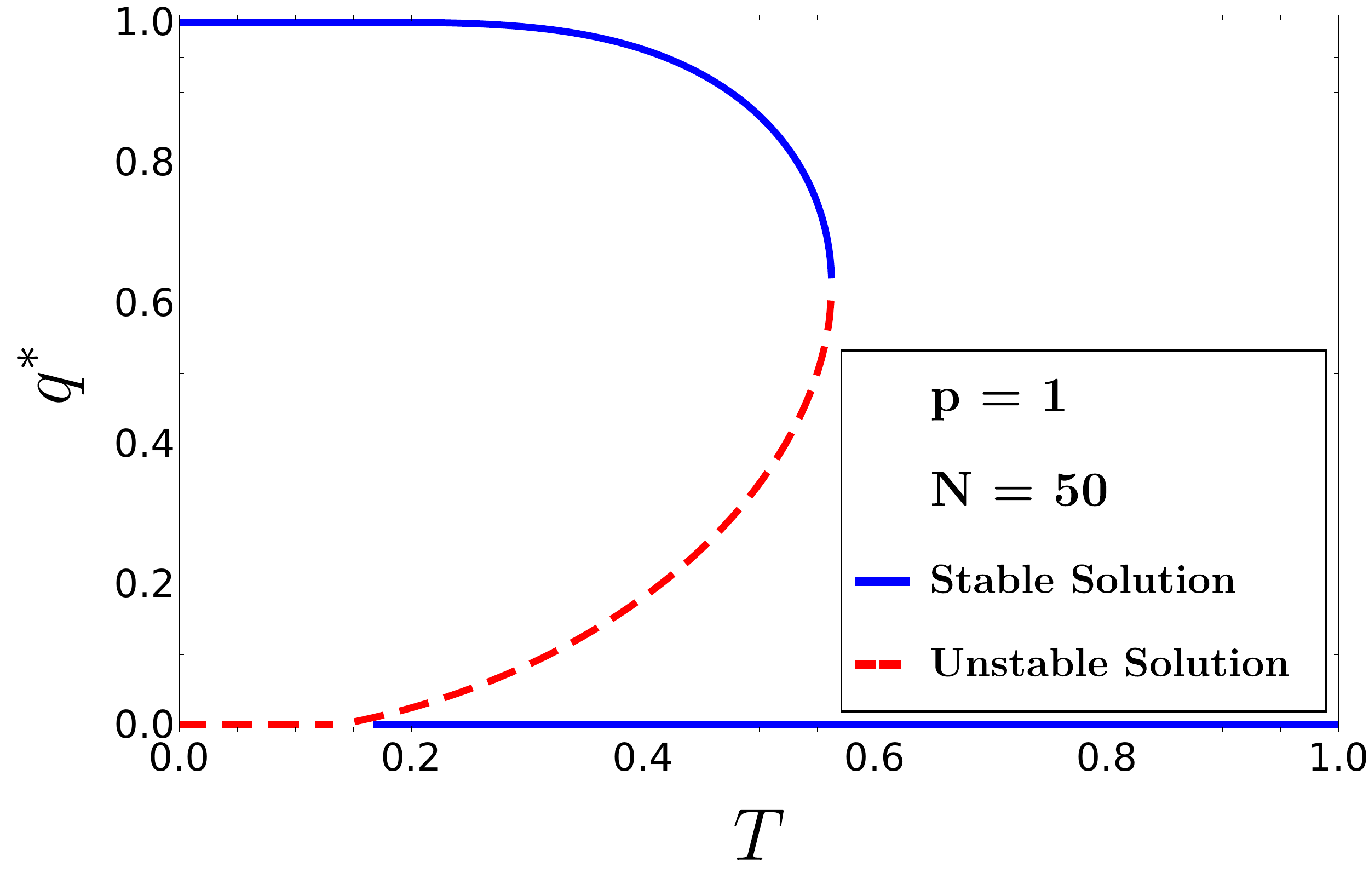}
		%\caption{Figure A}
		%\label{fig:a}
	\end{subfigure}
	\hfill
	\begin{subfigure}{0.48\linewidth}
		\includegraphics[width=\linewidth, height=0.6\linewidth]{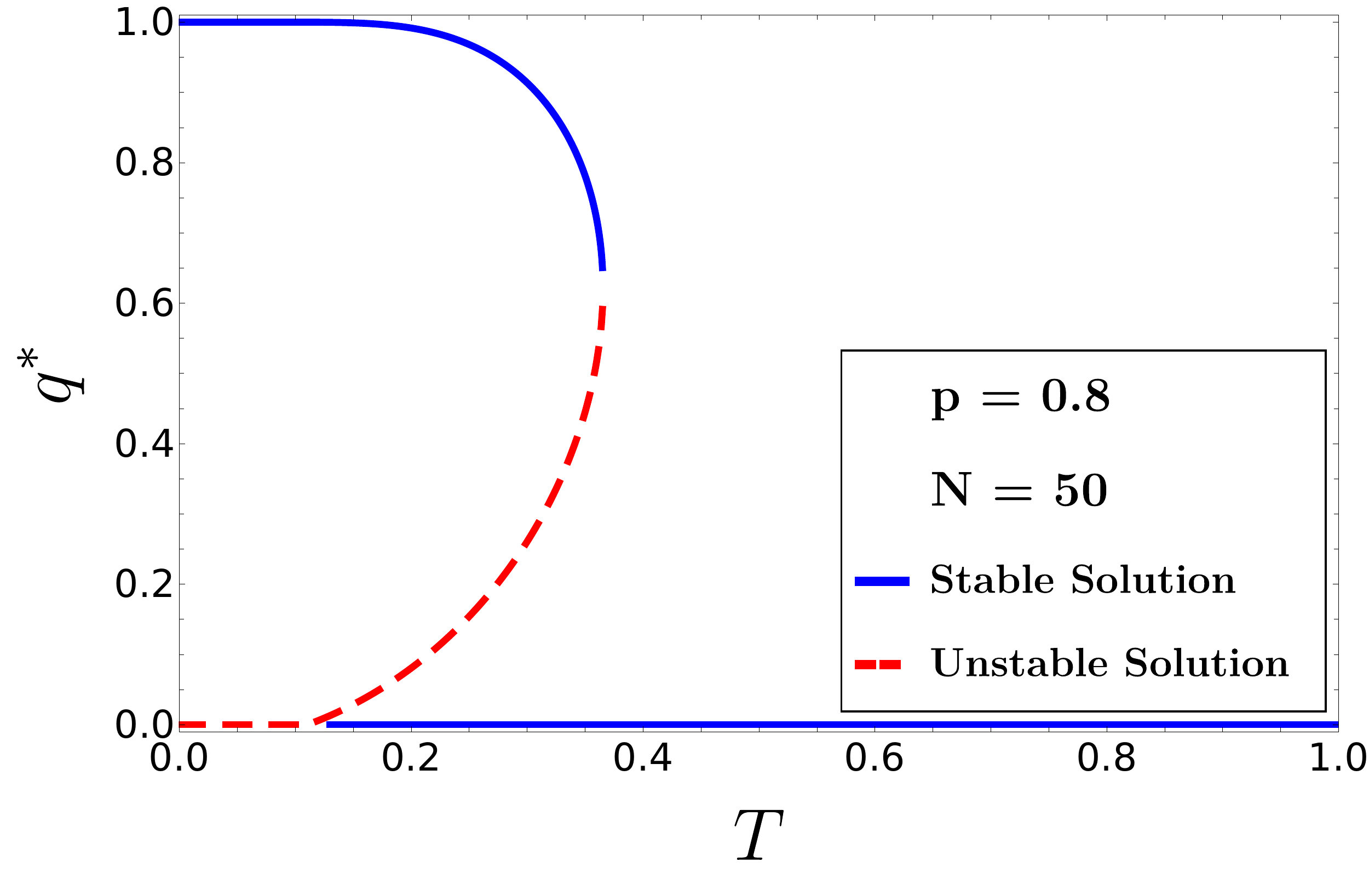}
		%\caption{Figure B}
		%\label{fig:b}
	\end{subfigure}
	
	\begin{subfigure}{0.48\linewidth}
		\includegraphics[width=\linewidth, height=0.6\linewidth]{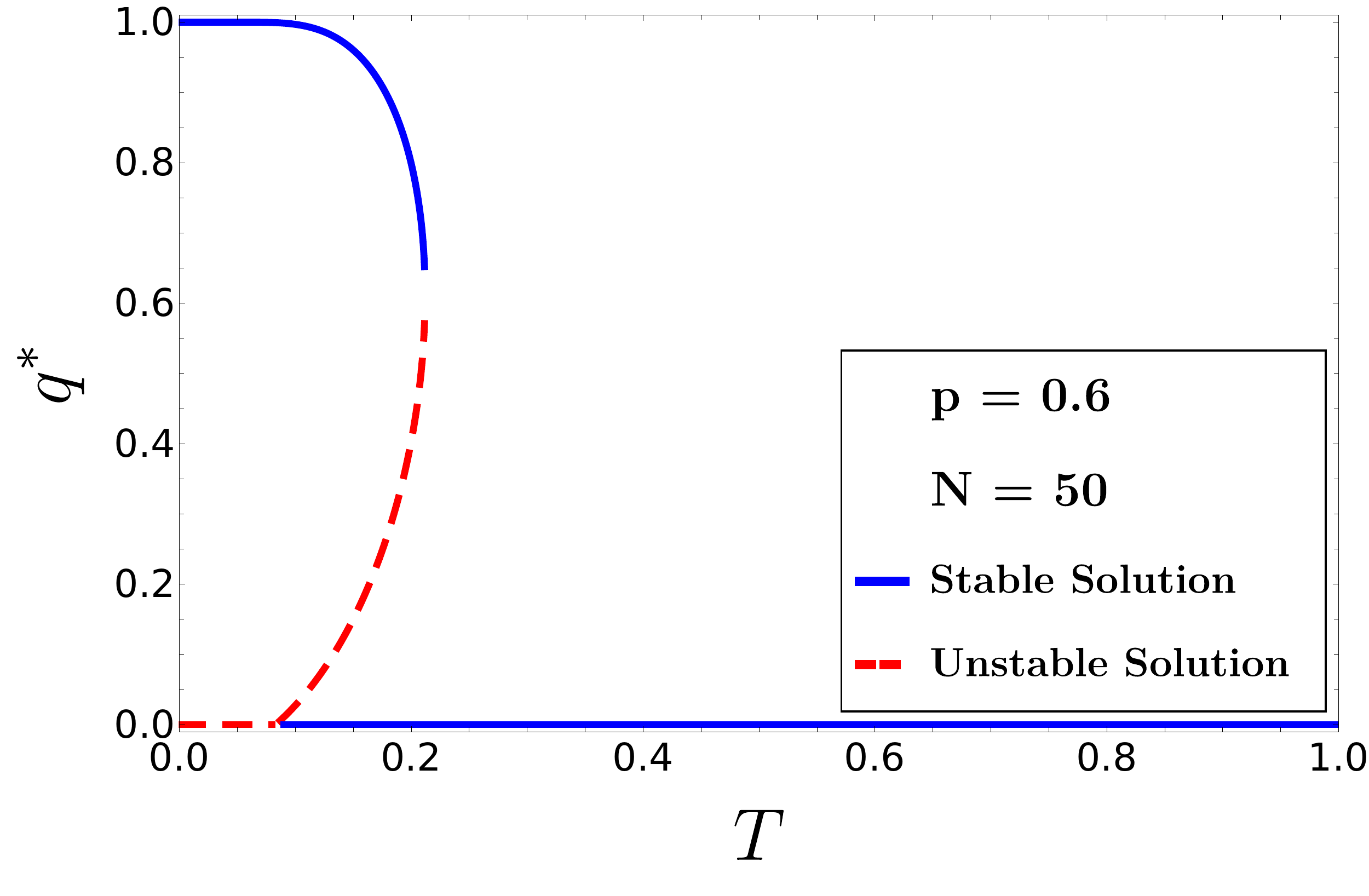}
		%\caption{Figure C}
		%\label{fig:c}
	\end{subfigure}
	\hfill
	\begin{subfigure}{0.48\linewidth}
		\includegraphics[width=\linewidth, height=0.6\linewidth]{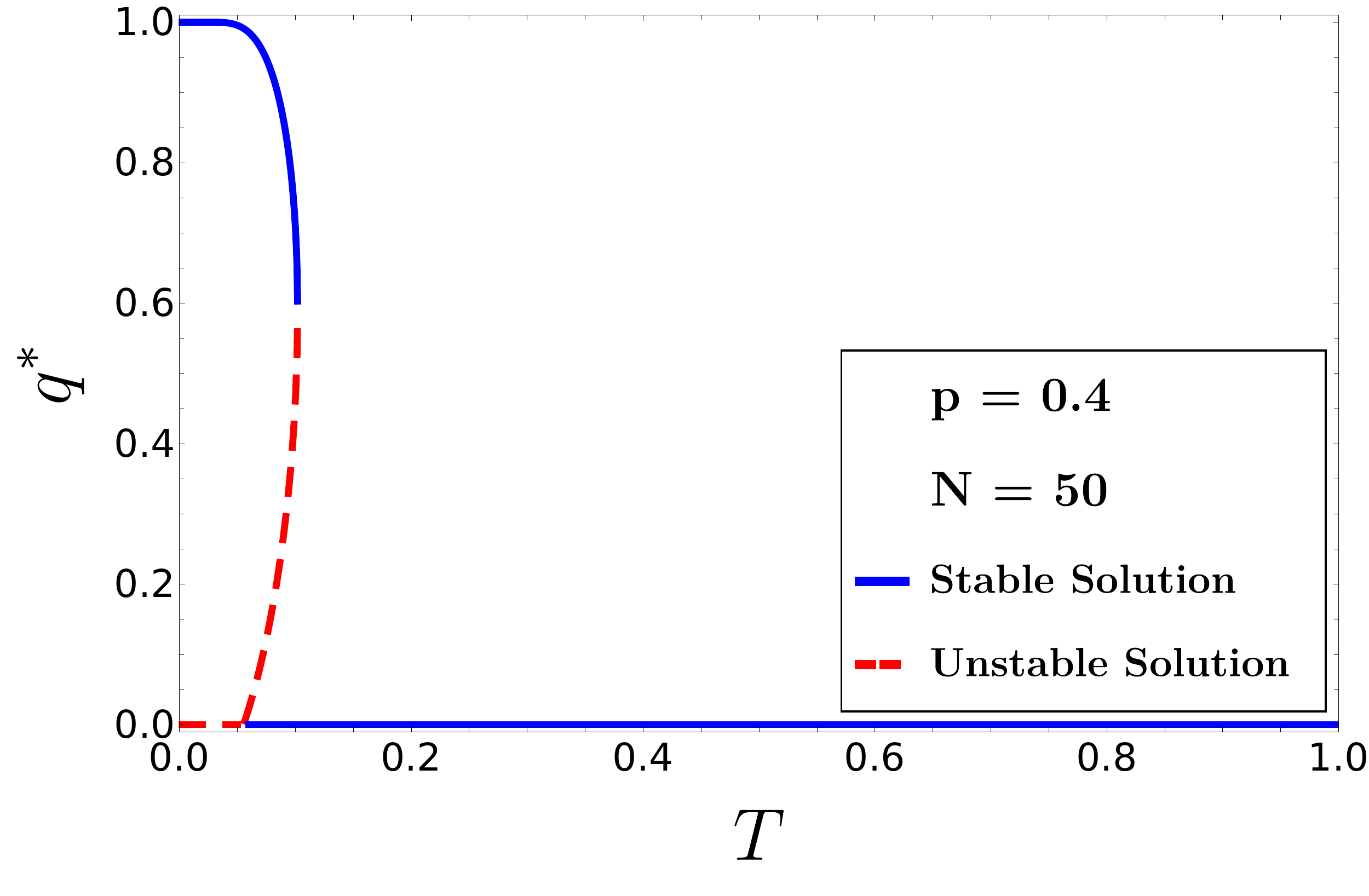}
		%\caption{Figure D}
		%\label{fig:d}
	\end{subfigure}
	\caption{\enquote {Blue-Sky} Bifurcation diagram versus temperature for four connection probability $p=1$, $p=.8$, $p=.6$ and $p=.4$ for an  Erd\H{o}s-R\'enyi Random graph of size $N=50$ model with Mean-Field approximation. For each $p$, three distinct regions are observed. For $T<T_{cold}$ the random graph is in balanced phase. For $T>T_{hot}$ it is completely in a random phase. For $T_{cold}<T<T_{hot}$ the random network experiences the coexistence phase.}\label{fig:1}
\end{figure*}

In Eq. \eqref{eq:8} we have used $\sum_{m=0}^{N-2}  mP(m)= (N-2)p^{2}$ which is the first moment of the binomial distribution. 
%From statistical mechanics we know that $\langle s_{ij} \rangle$ is the first derivation of $ [ \textbf{F}]$ respect to field coupled to $s_{ij}$, so we have:
For $\langle s_{ij} \rangle$ we have:
\begin{equation*} 
\begin{aligned}
\langle s_{ij} \rangle &=-\frac{\partial [\textbf{F}]_{c}}{\partial h_{ij}} {\Bigg|}_{h_{ij}=0}= -\frac{\partial [\textbf{F}^{\prime}]_{c^{\prime}}}{\partial h_{ij}}{\Bigg|}_{h_{ij}=0}\\
&+\frac{1}{\beta}  \frac{\partial }{\partial h_{ij}}  \ln\bigg(\cosh \big( \frac{\beta }{N} (N-2)p^{2}  {\big\langle ss \big \rangle}_{Z'} +\beta h_{ij} \big)\bigg) {\Bigg|}_{h_{ij}=0}\\
&=0+  \tanh \big( \frac{\beta }{N} (N-2)p^{2}  {\big\langle ss \big \rangle}_{Z'} +\beta h_{ij} \big) {\Bigg|}_{h_{ij}=0}
\end{aligned}
\end{equation*}
%%%%%%%%%%%%%%%%%%%%%%%%%%%%%%%%%%%%%%%%%%%%%%%%%%%%%%%%%%%%%%%%%%%%%%%%%%%%%%%%%%%%%%%%%%%%%
So, the average of signed edges, $\langle s \rangle$, is:
\begin{eqnarray}\label{eq:9}
\langle s \rangle = \tanh \big( \beta p^{2} \frac{N-2}{N} {\big\langle ss \big \rangle} \big)
\end{eqnarray} 
%As Eq.\eqref{eq:9} indicates $\langle s\rangle$ is a function of the ensemble average of two-stars, $\langle ss\rangle$. To calculate mean of two-stars which lead to a self-consistent equation for this quantity. For deriving $\langle ss \rangle$, we need a same method as we did in calculating $\langle s\rangle$. Following a similar method, we write the partition function while we separate the two-body term Hamiltonian $\textbf{H}_{jk,ki}$. Regarding the two-body term Hamiltonian, for partition function into $\textbf{H}_{jk,ki}$ and 

As Eq.\eqref{eq:9} indicates $\langle s\rangle$ is a function of the ensemble average of two-stars, $\langle ss\rangle$. To calculate $\langle ss\rangle$, we employ a similar method, where we split the Hamiltonian into $\textbf{H}_{jk,ki}$ and ${\textbf{H}}^{\prime \prime}$. The first term comprising the terms including at least one of the ties  $\{i,k\}$ and 
$\{k,j\}$, (Eq. \ref{eq:11}) and the second, comprising the rest.

\begin{equation}\label{eq:11}
\begin{aligned}
{\textbf{H}}_{ik, kj}&=-\frac{1}{N}s_{ik} \sum_{l\neq i,j,k} s_{il}s_{lk} e_{il}e_{lk} -\frac{1}{N}s_{kj}  \sum_{l\neq i,j,k} s_{kl}s_{lj} e_{kl}e_{lj}\\
&-\frac{1}{N} s_{ik}s_{kj}s_{ij}-s_{ik}s_{kj} h_{ik,kj},
\end{aligned}
\end{equation}

Where, $h_{ik,kj}$ is considered as a external field on two-star $s_{ik}s_{kj}$. The partition function can be written as follows:

\begin{equation}\label{eq:10}
\begin{aligned}
\textbf{Z}&=\sum_{\{\textbf{s}\}} e^{-\beta \textbf{H}}= \sum_{\{\textbf{s}\}} e^{-\beta ({\textbf{H}}_{ik,kj}+{\textbf{H}}^{\prime \prime})}\\
&={\textbf{Z}}^{\prime \prime} \sum_{s\neq s_{ik},s_{kj}} \frac{e^{-\beta {{\textbf{H}}}^{\prime \prime}}}{{\textbf{Z}}^{\prime \prime}} \sum_{s_{ik}=\pm 1} \sum_{s_{kj}=\pm 1} e^{-\beta {\textbf{H}}_{ik,kj}}\\
&={\textbf{Z}}^{\prime \prime} {\bigg \langle   \sum_{s_{ik}=\pm 1} \sum_{s_{kj}=\pm 1}  e^{-\beta {\textbf{H}}_{ik,kj}}\bigg \rangle }_{Z^{\prime \prime}}
\end{aligned}
\end{equation}
%$H_{ik,kj}$ is a part of Hamiltonian which contains ties $\{i,k\}$ or $\{k,j\}$ or both of them simultaneously, So for existing $e_{jk}=e_{ki}=1$,  $H_{ik,kj}$ is written as bellow:

Substituting \eqref{eq:11} in \eqref{eq:10} we have:
\begin{equation}\label{eq:12}
\begin{aligned}
\textbf{Z}&=\textbf{Z}^{\prime \prime} \bigg ( e^{\frac{\beta}{N}\big(m_{1} \langle s_{il}s_{lk}\rangle_{Z^{\prime \prime}} +m_{2}\langle s_{kl}s_{lj}\rangle_{Z^{\prime \prime}}+\langle s_{ij}\rangle_{Z^{\prime \prime}}\big)+\beta h_{ik,kj}  } \\
&+ e^{\frac{\beta}{N}\big(m_{1} \langle s_{il}s_{lk}\rangle_{Z^{\prime \prime}} -m_{2}\langle s_{kl}s_{lj}\rangle_{Z^{\prime \prime}}- \langle s_{ij}\rangle_{Z^{\prime \prime}}\big) -\beta h_{ik,kj}  }\\
&+e^{\frac{\beta}{N} \big(-m_{1} \langle s_{il}s_{lk}\rangle_{Z^{\prime \prime}} +m_{2} \langle s_{kl}s_{lj}\rangle_{Z^{\prime \prime}} -\langle s_{ij}\rangle_{Z^{\prime \prime}}\big ) -\beta h_{ik,kj} }\\
&+e^{\frac{\beta}{N}\big(-m_{1} \langle s_{il}s_{lk}\rangle_{Z^{\prime \prime}} -m_{2}\langle s_{kl}s_{lj} \rangle_{Z^{\prime \prime}}+ \langle s_{ij}\rangle_{Z^{\prime \prime}}\big) +\beta h_{ik,kj}  } \bigg),
\end{aligned}
\end{equation}

%Where $m_{1}$ is the number of two-stars established on tie $\{i,k\}$ not including $j$ and $m_{2}$ is the number of two-stars established on tie $\{k,j\}$ that do not include node $i$.

Where $m_{1}$ and $m_{2}$ are the number of triangles established on ties $\{i,k\}$ and $\{k,j\}$, respectively not including nodes $j$ and $i$.

%\Sina{I suggest using the signs $m_{ \cancel{j} }$ and $m_{ \cancel{k} }$, instead of $m_{1}$ and $m_{2}$ }

The homogeneity of Erd\H{o}s-R\'enyi random graph allows us to assume $m_{1}=m_{2}$. From statistical mechanics we have \cite{Amirhossein}:
\begin{eqnarray}\label{eq:122}
\langle s_{ik}s_{kj} \rangle &=-\frac{\partial \big[\textbf{F}\big]_{c}}{\partial h_{ik,kj}} {\Bigg|}_{h_{ik,kj}=0}
\end{eqnarray}
%Ensemble average of the two-stars is the first derivation of configuration free energy respect to field $h_{ik,kj}$. 
So for mean of two-stars we have:
%So for $\langle ss \rangle $ we have:
\begin{equation}\label{eq:13}
\begin{aligned}
\langle s s\rangle =
\frac{\bigg( e^{\beta(2p^2 \frac{N-3}{N}\langle ss\rangle )} - 2e^{{\beta}(-2\frac{\langle s \rangle}{N} )} + e^{\beta(-2 p^2 \frac{N-3}{N} \langle ss\rangle  )} \bigg)}{\bigg( e^{\beta(2 \frac{N-3}{N} p^2 \langle ss\rangle )} + 2e^{{\beta}(-2\frac{\langle s \rangle}{N} )} + e^{\beta(-2 p^2 \frac{N-3}{N}\langle ss\rangle  )} \bigg)}
\end{aligned}
\end{equation}
Details of this calculation are given in Appendix. \ref{appendix:c}.
%%%%%%%%%%%%%%%%%%%%%%%%%%%%%%%%%%%%%%%%%%%%%%%%%%%%%%%%%%%%%%%%%%%%%%%%%%%%%%%%%%%%%%%%%%%%
By replacing Eq.\eqref{eq:9} in Eq.\eqref{eq:13} we reach a self-consistent equation that yields the ensemble average of the two-stars for any temperature $T$, based on the connection probability $p$ and network size $N$. The intersections of Eq.\eqref{eq:13} yield its fixed points. We illustrate this result for different $p$ values, in Appendix \ref{appendix:a}.
%\begin{eqnarray}
%<s_{ik} s_{kj}>= -\frac{\partial [\textbf{F}]}{\partial h_{ik, kj}} {\Bigg|}_{h_{ik,kj}=0}=\\
%{\bigg< \sum_{U=0}^{N-3} \sum_{V=0}^{N-3} P(V)P(W) a_{ik}a_{kj}<Z>_{Z^{\prime}} {\Bigg|}_{U,V}  %\bigg>}_{C^{\prime}}=\\
%{\bigg< \sum_{U=0}^{N-3} \sum_{V=0}^{N-3} P(V)P(W) a_{ik}a_{kj}{\bigg <{e^{\beta( V s_{il} s_{lk} +h_{ik} %a_{ik}+ W s_{kl}s_{lj} + h_{kj}a_{kj} +h_{ij}s_{ij} )} + e^{\beta( V s_{il} s_{lk} +h_{ik} a_{ik}- W %s_{kl}s_{lj} - h_{kj}a_{kj} -h_{ij}s_{ij})}\\
% &{+e^{\beta( -V s_{il} s_{lk} -h_{ik} a_{ik}+ W s_{kl}s_{lj} +h_{kj}a_{kj} -h_{ij}s_{ij})}+ e^{\beta( -V s_{il} s_{lk} -h_{ik} a_{ik}- W s_{kl}s_{lj} -h_{kj}a_{kj} +h_{ij}s_{ij})} }
% }\bigg >}_{Z^{\prime}} {\Bigg|}_{U,V}  \bigg>}_{C^{\prime}}
%\end{eqnarray}
%\end{widetext}
%%%%%%%%%%%%%%%%%%%%%%%%%%%%%%%%%%%%%%%%%%%%%%%%%%%%%%%%%%%%%%%%%%%%%%%%%%%%%%
%\begin{widetext}
Fig.\ref{fig:1} depicts the bifurcation diagram versus temperature for different connection probabilities in an Erd\H{o}s-R\'enyi graph of size $N$. As it is shown in Fig.\ref{fig:1}, the bifurcation is of the \enquote{blue-sky} type, which is the characteristic of a discontinuous phase transition which leads to three distinct regions:
\begin{itemize}
	\item For $T>T_{hot}$ or high-temperature regime: There exists one stable fixed point with $q^{*}=0$. From an intuitive point of view, $q^{*}=0$ refers to the random phase, in the sense that high thermal fluctuations prevent the formation of any order in the system.
	
	\item For $T_{cold}<T<T_{hot}$ or coexistence region: There exist three fixed points of which two are stable $(q_{1}^{*}=0, q_{2}^{*})$ and the other one is unstable. Since we have two stable fixed points in this region the system experiences the coexistence of both random and balanced phases. The hysteresis loop is obtained due to the coexistence of balanced and random phases within a specified temperature range.
	%\Sina{I don't get it, cite and explain the hysteresis loop.}
	\item For $T<T_{cold}$ or low-temperature regime: There exist two fixed points with $q_{1}^{*}=0$ and $q_{2}^{*}=1$. The first one is an unstable fixed point and the second one is stable. From an intuitive point of view, when the temperature is low enough, a long-ranged order is formed and the system is in a total balance phase. In other words, ties are frozen because of a strong field $q^{*}=1$. 
	
\end{itemize}

As it is shown for \enquote{blue-sky} bifurcation diagram in Fig.\ref{fig:1} the cold critical temperature, $T_{cold}$ is where the coexistence region starts appearing. To derive this point analytically, we need to take the first derivative of the self-consistent Eq.\ref{eq:13} with respect to $q$:
\begin{figure*}[!t]
	\begin{subfigure}{0.48\linewidth}
		\includegraphics[width=\linewidth, height=0.8\linewidth]{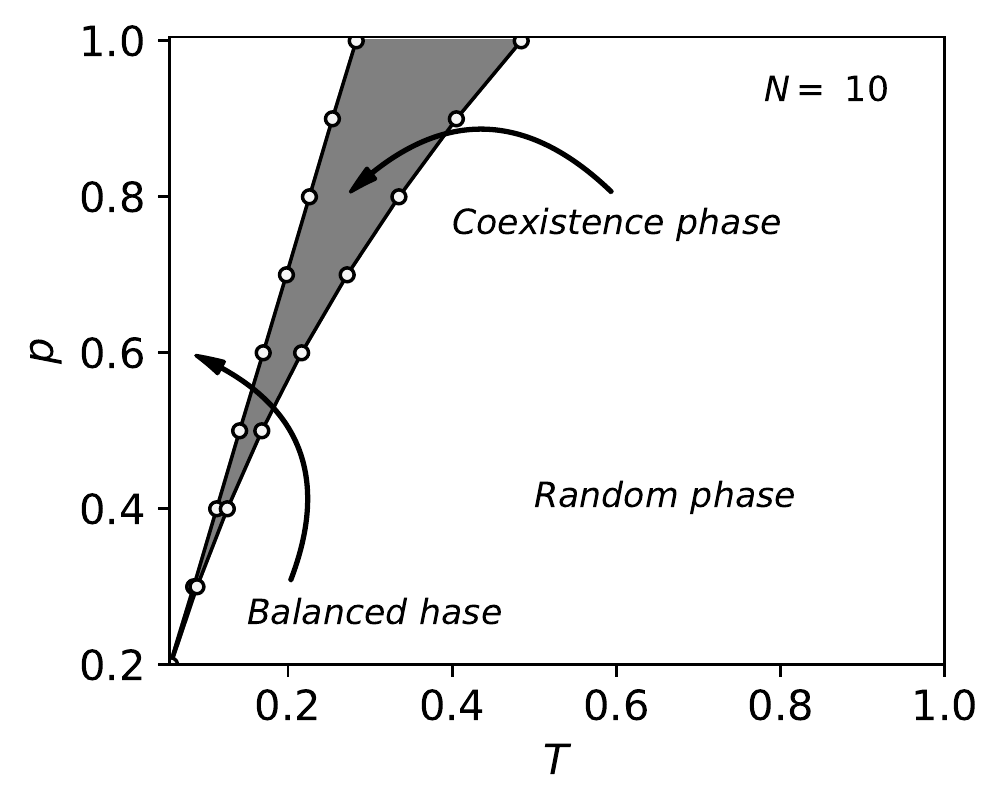}
		%\caption{Figure A}
		%\label{fig:a}
	\end{subfigure}
	\hfill
	\begin{subfigure}{0.48\linewidth}
		\includegraphics[width=\linewidth, height=0.8\linewidth]{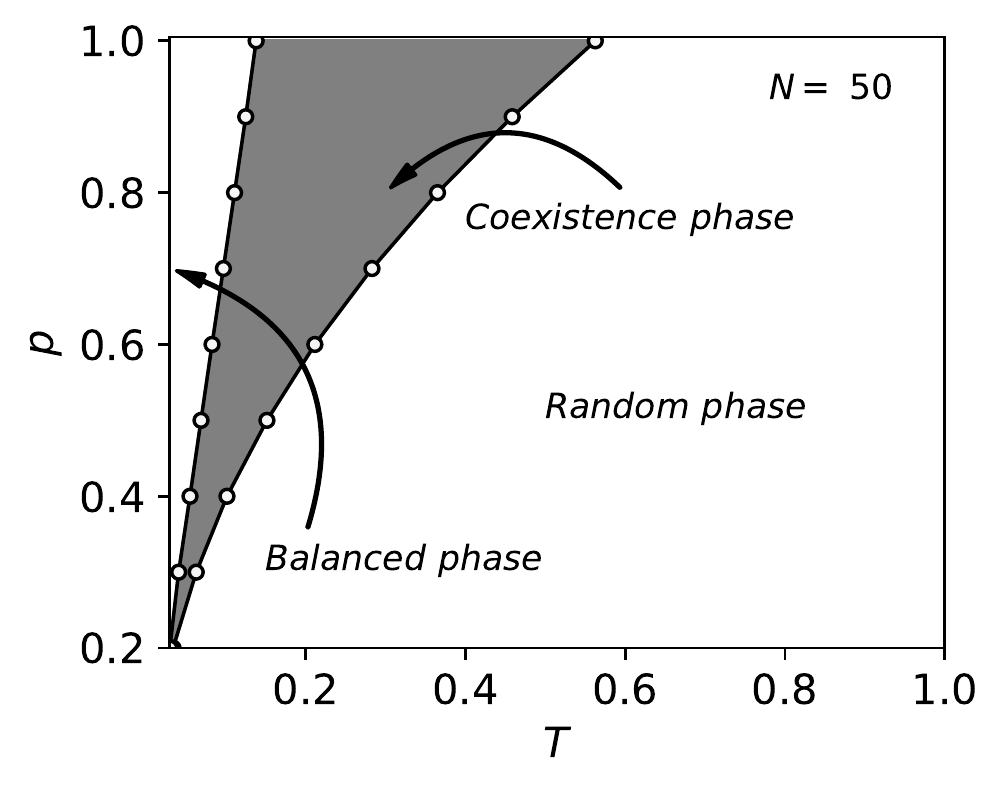}
		%\caption{Figure B}
		%\label{fig:b}
	\end{subfigure}
	\caption{The phase diagram in $(T,p)$ space for an  Erd\H{o}s-R\'enyi random graph of size $N=10$ and $N=50$ respectively. As it is indicated the shaded region represents the coexistence phase in which, random and balanced phases coexist.}\label{fig:2}
\end{figure*}
%%%%%%%%%%%%%%%%%%%%%%%%%%%%%%%%%%%%%%%%%%%%%%%%%%%%%%%%%%%%%%%%%%%%%%%%%%%%%%%%%%%%%%%%%%%%%%
As we know from stability analysis, the fixed points of a self-consistent equation $f(q^{*})=q^{*}$ are stable if $f^{\prime} (q^{*})<1$ and unstable if $f^{\prime} (q^{*})>1$.
Thus, $q^{*}=0$ is stable when $f^{\prime} (q^{*}=0)={\beta}^{2} {p}^{2} (\frac{N-2}{N^2})<1$. Therefore we have:
\begin{equation}\label{eq:145}
T_{cold}=p \sqrt{\frac{N-2}{N^{2}}}
\end{equation}
Hence, the cold critical temperature has a linear dependency on the connection probability $p$. Furthermore, in the limit of large $N$ values, the cold critical temperature converges to zero. This leads to vanishing the purely balanced phase even for low temperatures.
%However, this does not occur in empirical (social) networks.
%We know that structural balance and consequently our model deals with triadic motifs in networks. On the other hand, we know that in an Erd\H{o}s-R\'enyi random graph with a desired connection probability, triads are distributed homogeneously in all parts of the graph. Now, if we look at an empirical network in a sub-graph scale and an Erd\H{o}s-R\'enyi random graph in a mesoscopic scale comparable with that sub-graph scale, the empirical network prone to have more triangles. But because of sparsity, different sub-graphs in an empirical networks do not have many connections with each other, so in the macro-scale the total number of triangles in an empirical networks is significantly low. 
%and an Erd\H{o}s-R\'enyi random graph with the same number of nodes and links, the empirical network prone to have more triadic motifs in its sub-graphs. So,  the number of triangles in a large empirical network in a Mesoscopic scale (in its sub-graphs) is comparable with an Erd\H{o}s-R\'enyi random graph with an intermediate size.
%We have conducted our calculations for an Erd\H{o}s-R\'enyi random graph of size $N=50$.
%we know that empirical networks (social network?) \Sina{Real or random?} are considerably sparse.

%\Sina{I have several issues with this section. First, the number of triangles should be compared conditioned on the equality of the number of edges, otherwise, of course the denser network has more triangles as well. Also, the term \enquote{real network} is vague, better to use empirical or social networks.}

In Fig.\ref{fig:2} we have illustrated the phase diagram of the ensemble average of the two-stars as a function of $p$ and $T$ to analyze the behavior of the coexistence region in the $(p, T)$ phase space.
Fig.\ref{fig:2} indicates the phase diagram in $(p,T)$ space for an  Erd\H{o}s-R\'enyi random graph with connection probability $p$ at temperature $T$ for network sizes $N=50$ and $N=10$.
As it is shown, the phase space is divided into three regions. The shaded region represents the $p$ and $T$ values leading to a bi-stability in the system, in which, both random and balanced phases coexist.

As it is illustrated, an increase in the size of the random graph enlarges the coexistence region and decreases $T_{cold}$.

In the next section, we verify our Mean-Field solution by a series of Monte-Carlo simulations.
\begin{figure*}[!t]
	\begin{subfigure}{0.48\linewidth}
		\includegraphics[width=\linewidth, height=0.7\linewidth]{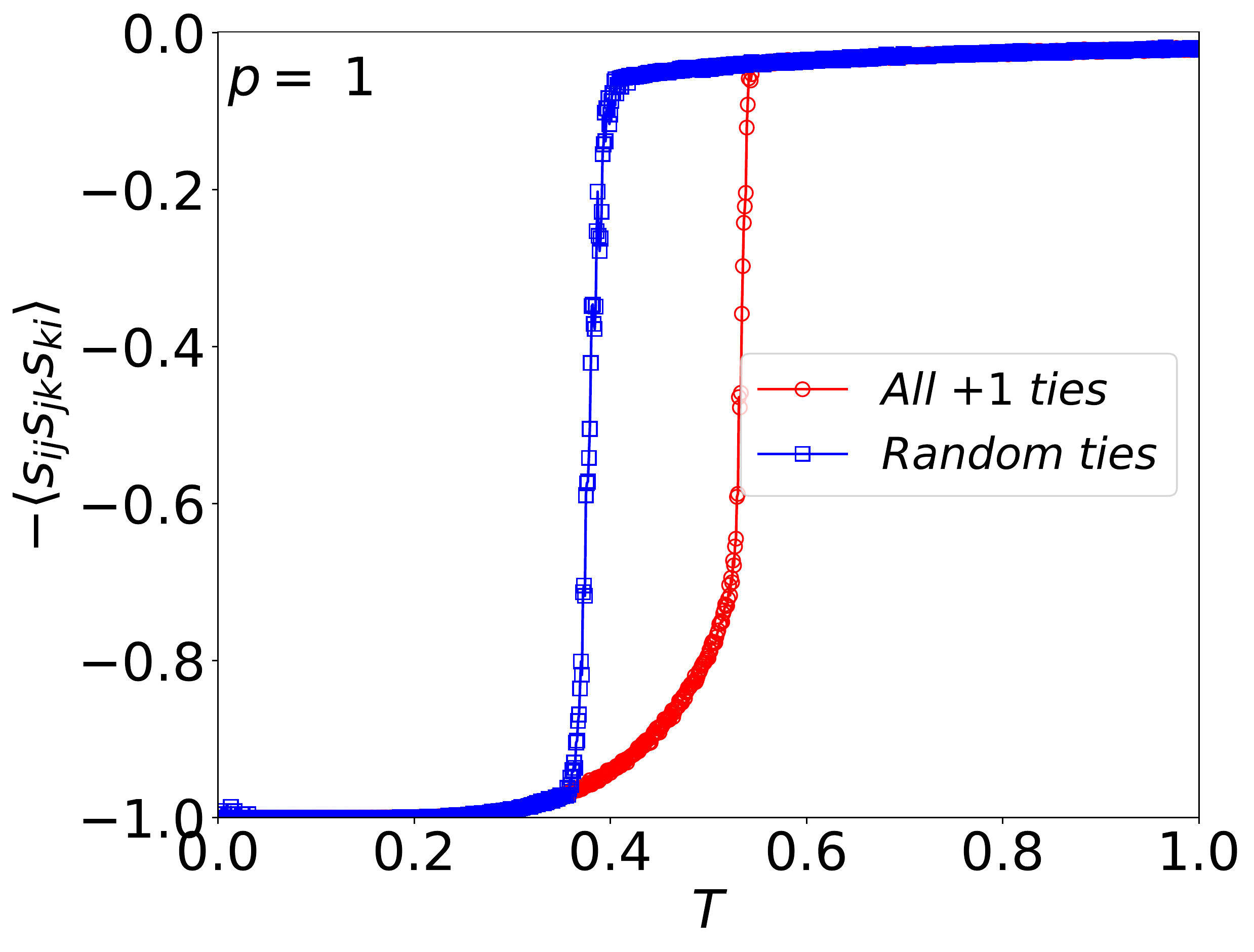}
		%\caption{Figure A}
		%\label{fig:a}
	\end{subfigure}
	\hfill
	\begin{subfigure}{0.48\linewidth}
		\includegraphics[width=\linewidth, height=0.7\linewidth]{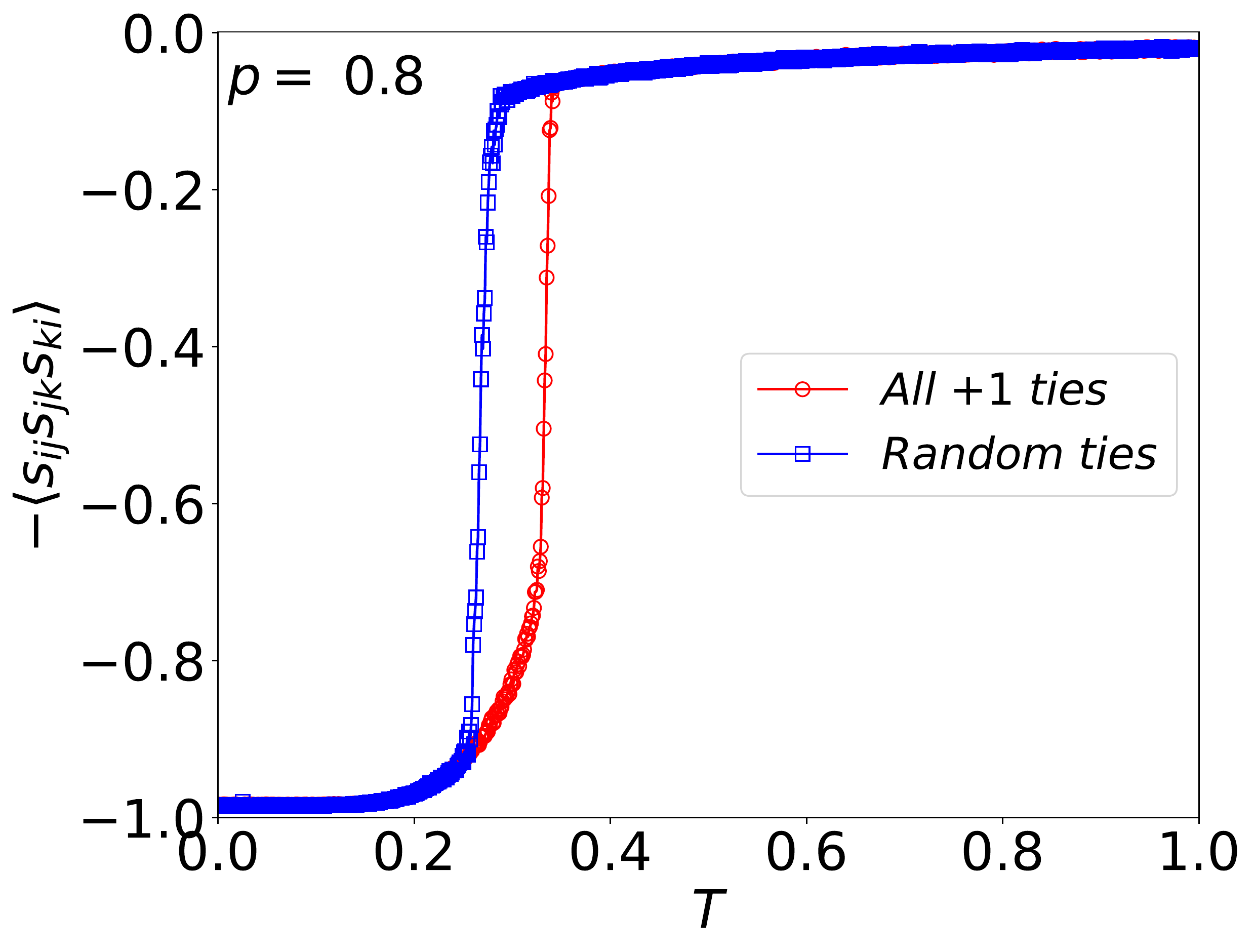}
		%\caption{Figure B}
		%\label{fig:b}
	\end{subfigure}
	
	\begin{subfigure}{0.48\linewidth}
		\includegraphics[width=\linewidth, height=0.7\linewidth]{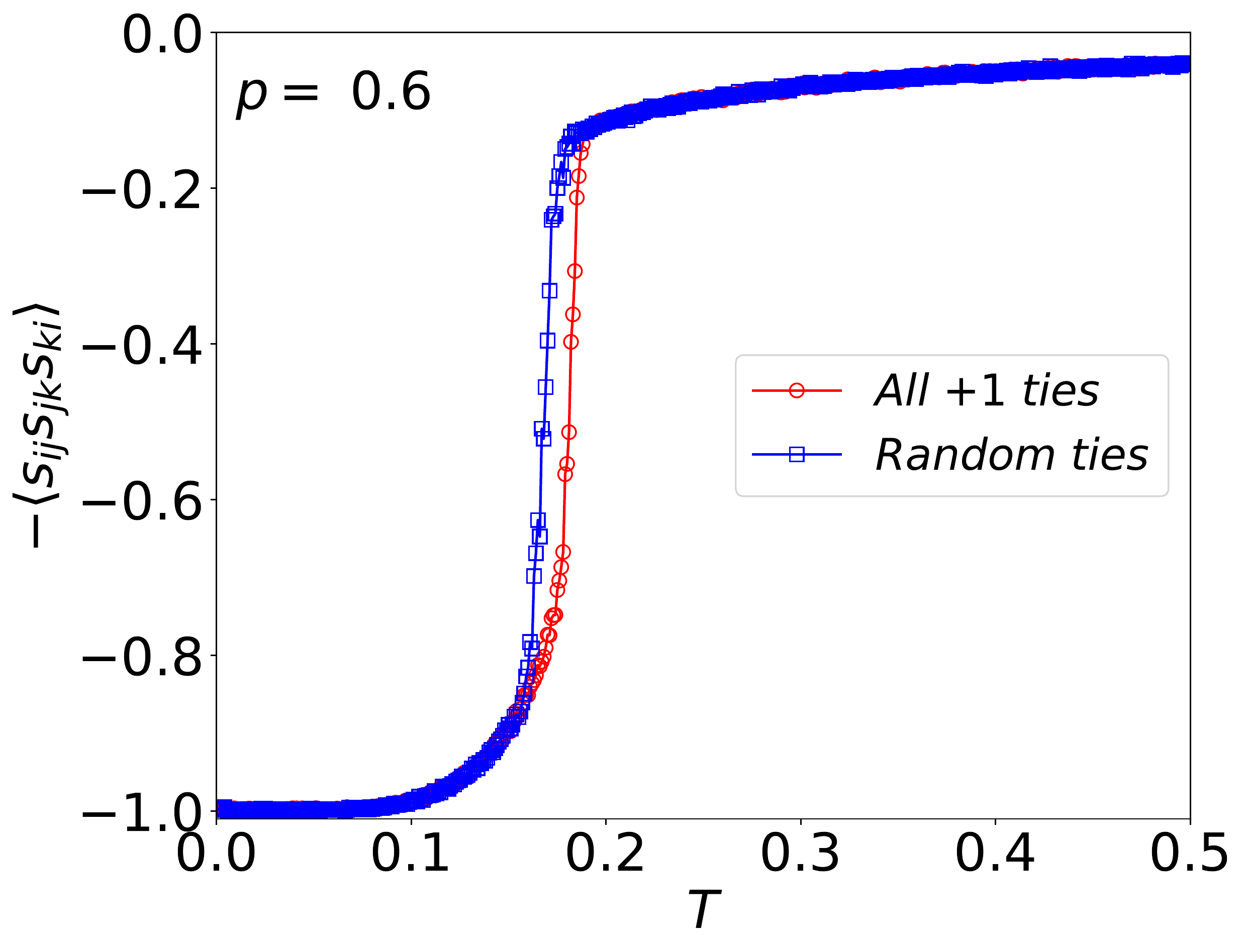}
		%\caption{Figure C}
		%\label{fig:c}
	\end{subfigure}
	\hfill
	\begin{subfigure}{0.48\linewidth}
		\includegraphics[width=\linewidth, height=0.7\linewidth]{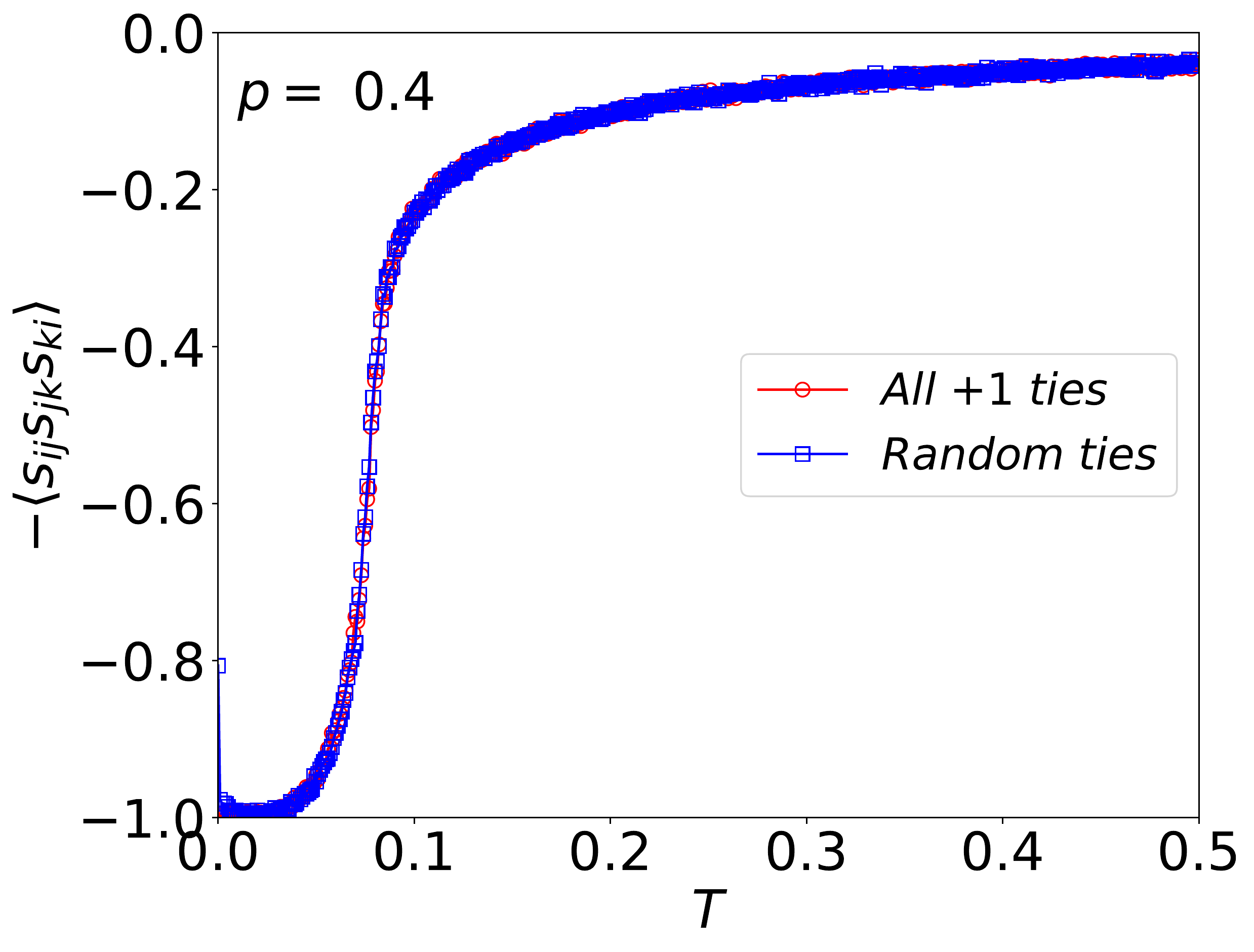}
		%\caption{Figure D}
		%\label{fig:d}
	\end{subfigure}
	\caption{The mean of Energy versus temperature for four connection probability $p=1$, $p=.8$, $p=.6$ and $p=.4$ in an Erd\H{o}s-R\'enyi Random graph of size $N=50$. As it is apparent, with increasing the network sparsity both $T_{hot}$ and $T_{cold}$ decrease, but the slope of decreasing $T_{hot}$ is more than the slope of decreasing $T_{cold}$. Therefore by reaching to an specific value of sparsity the coexistence region totally vanishes}\label{fig:3}
\end{figure*}
%%%%%%%%%%%%%%%%%%%%%%%%%%%%%%%%%%%%%%%%%%%%%%%%%%%%%%%%%%%%%%%%%%%%%%%%%%%%%%%%%%%%%%%%%%%%%%%%%%%%%%%%%%%%%
\section{Simulation}
We generate $500$ realizations of an Erd\H{o}s-R\'enyi random graph model with size $N=50$ and connection probability $p$.
Each tie is initially in the $(+)$ state with probability $\alpha$ and in the $(-)$ state with probability $1-\alpha$.
We consider two different initial conditions:
(i) $\alpha = 1$ (All $+$).
(ii) $\alpha = 0.5$ (Random signs).
%We consider two different initial conditions:
%(i) All ties in the $(+)$ state.
%(ii) Ties in the $(+)$ state with probability $\alpha = 0.5$.
To reach a stationary state of the system, we apply a Metropolis-Hastings algorithm on the tie signs. The algorithm is as follows:
% with respect to our Hamiltonian (see Eq. \eqref{eq:1}) and temperature $T$.

\begin{itemize}
	\item We choose a random tie and consider flipping its sign. If the energy variation is negative, i.e., $\Delta \textbf{E}=\textbf{E}_{f} - \textbf{E}_{i}<0$, the flip is accepted. Where $E_i$ and $E_f$ indicate the energy of the system before and after the flip.
	
	\item If the energy variation is semi-positive, i.e.,  $\Delta \textbf{E}=\textbf{E}_{f} - \textbf{E}_{i}\geq 0$, the flip is accepted with the Boltzmann probability $e^{-\nicefrac{\Delta \textbf{E}}{\textbf{T}}}$.

	\item This procedure continues until the system reaches the stationary state.

\end{itemize}
%For each realization, we consider $N^{3}$ Monte-Carlo steps and then we start sampling.

Fig. \ref{fig:3} illustrates the average energy, $\langle \textbf{E} \rangle=-\langle s_{ij} s_{jk} s_{ki}\rangle$, versus temperature for different connection probabilities.
We observe two curves, each corresponding to an initial condition: the curve obtained for the random initial condition has a critical point at $T=T_{cold}$, and the curve obtained for the all $+1$ initial condition has a critical point at $T=T_{hot}$. Therefore, simulation results correctly capture the two critical temperatures $T_{cold}$ and $T_{hot}$.

Furthermore, By decreasing the connection probability $p$, the coexistence region becomes narrower until in $p \approx 0.4$ it completely vanishes.
In addition, the hot critical temperature, $T_{hot}$, is in good agreement with the Mean-Field solution.

On the other hand, since the basin of attraction of $q^{*}=0$ in the coexistence region is very narrow, it is practically challenging to capture $T_{cold}$ via the simulation for all temperatures.
%%%%%%%%%%%%%%%%%%%%%%%%%%%%%%%%%%%%%%%%%%%%%%%%%%%%%%%%%%%%%%%%%%%%%%%%%%%%%%%%%%%%%%%%%%%%%%%%%%%%%%%
%%%%%%%%%%%%%%%%%%%%%%%%%%%%%%%%%%%%%%%%%%%%%%%%%%%%%%%%%%%%%%%%%%%%%%%%%%%%%%%%%%%%%%%%
\section*{Conclusion}
The idea of standard balance on empirical data obtained from real singed networks has brought the opportunity to capture interesting phenomena in political networks, psychology, international relations, and ecology. However, not much has been done so far, to investigate signed random networks with an analytical approach. 
We know that real random signed networks are indeed sparse. On the other hand, the idea of completely eliminating tension in triad relationships in signed networks inspired by structural balance theory does not come true for many signed networks. For instance, in social networks, agents may tend to change their relations with the other agents, even though these changes are not in favor of total tension reduction. Thus, the concept of social temperature has been introduced to capture a different level of tension tolerance  \cite{Amirhossein}. Therefore, there is a competition between agents' random behavior (social temperature) and the tendency to reach the state of balance. Considering these two issues, we have proposed a model that defines the structural balance on Erd\H{o}s-R\'enyi networks in the presence of temperature within a theoretical framework.

We have solved the model with a Mean-Field approach. We have detected a discontinuous phase transition with temperature. We have captured two temperatures, $T_{cold}$ and $T_{hot}$ which leads to three distinct regions: 
%which are the characteristic temperatures of this abrupt phase transition.
For $T<T_{cold}$, the system is in a completely balanced phase. On the other hand, for $T>T_{hot}$, the system can not reach the balanced phase. for $T_{cold}<T<T_{hot}$, the system demonstrates bi-stability with two stable fixed points $q_{1}^{*} =0$ and $q_{2}^{*}$ respectively denoting the random and balanced phases. 
%Thus, even for low temperatures, the system state can converge to the basin of attraction of $q^{*}=0$ and be trapped in this nontrivial fixed point.
%We have derived $T_{cold}$ analytically. Our model predicts that for large random networks, the cold temperature tends to zero. It means that for a large enough random network the model predicts that even for low enough temperatures $(T \approx 0)$ the system may be trapped in a random phase, while we expect a purely balanced phase for $T \approx 0$.
Regarding the bi-stability region, the system experiences a hysteresis phenomenon. It means that depending on the initial condition, the system meets one of the two curves that surround the coexistence region. Therefore, we do not need to over-cool the system to enter the bipolar state. The system enters the bipolar state even for non-zero temperatures. Another outcome of the model is that: the more sparse the network, the closer the $T_{cold}$ and $T_{hot}$ become. Hence, for sparser networks, it is easier to get out of a bipolar state in lower temperatures. 
Solving $T_{cold}$ analytically, we derive that for large E-R networks $T_{cold}$ converges to 0. Counter-intuitively, this indicates that even for $T \approx 0$ the system can be in a random state.

%However, the real random networks do not meet this condition, because in our model the triangles are homogeneously distributed all over parts of the networks. But in empirical networks, the number of triangles in the sub-sets of a random graph is relatively low.
%\Sina{Again I don't understand whether the distribution or the number of triangles are being discussed.}

We spanned the phase space $(T,p)$ to analyze the coexistence region. We observed that by reaching a specific value of connection probability the coexistence region vanishes. We have conducted a series of Monte-Carlo simulations to verify the result we obtained by the Mean-Field approach. There is a good agreement for the behavior of the hot temperature between the two approaches. The minor discrepancy for the cold temperature
in the two approaches is due to the fact that the basin of attraction of $q^{*}=0$ is narrow and hard to completely capture in a Monte-Carlo simulation.
%%%%%%%%%%%%%%%%%%%%%%%%%%%%%%%%%%%%%%%%%%%%%%%%%%%%%%%%%%%%%%%%%%%%%%%%%%%%%%%%%%%%%%%%%%%%%%%%%
\begin{widetext}
\begin{appendices}
\section{Configurational free energy: one-body Hamiltonian approach }
\label{appendix:b}
For calculating the \textit{configurational free energy}, we have to take the configurational average of the \textit{quenched partition function} over all possible random graph configurations. Therefore we have:
\begin{equation}\label{eq:14}
\begin{aligned}
{[\textbf{F}]_{c}}&= -\frac{1}{\beta} [\ln {\textbf{Z}} ]_{c}=-\frac{1}{\beta} [\ln ( {\textbf{Z}_{ij}}\textbf{Z}^{\prime}) ]_{c}= -\frac{1}{\beta} [\ln \textbf{Z}^{\prime} +\ln \textbf{Z}_{ij}]_{c}=-\frac{1}{\beta}[\ln \textbf{Z}^{\prime}]_{c}-\frac{1}{\beta}[\ln \textbf{Z}_{ij}]_{c}\\
&=-\frac{1}{\beta}[[\ln \textbf{Z}^{\prime}]_{c_{ij}}]_{c^{\prime}} 
-\frac{1}{\beta}[[\ln \textbf{Z}_{ij}]_{c^{\prime}}]_{c_{ij}} \\
&=-\frac{1}{\beta}[\ln \textbf{Z}^{\prime}]_{c^{\prime}} 
-\frac{1}{\beta}[\ln \textbf{Z}_{ij}]_{c_{ij}}\\
&=-\frac{1}{\beta} \bigg( [\textbf{F}^{\prime}]_{c^{\prime}} +  [\textbf{F}_{ij}]_{c_{ij}}\bigg).
% -\frac{1}{\beta} \sum_{\{c_{ij}\}} \sum_{\{c^{\prime}\}} \ln \bigg (\textbf{Z}_{ij} \textbf{Z}^{\prime} \bigg)\\
% &= -\frac{1}{\beta} \sum_{\{c_{ij}\}} \sum_{\{c^{\prime}\}} \ln \textbf{Z}^{\prime} -\frac{1}{\beta} \sum_{\{c^{\prime}\}} \sum_{\{c_{ij}\}} \ln \textbf{Z}_{ij}
\end{aligned}
\end{equation}
In Eq. \ref{eq:14}, the bracket $[\hspace{2mm}]_{c}$ indicates the configurational average of partition function over all possible random graph configurations. As we explained in the model section, we can decompose the partition function to two parts: $\textbf{Z}_{ij}$ and $\textbf{Z}^{\prime}$ that existing tie $\{i,j\}$ does and does not contribute to. Therefore we can write the configurational free energy as the sum of two terms: the first term of Eq. \eqref{eq:14}, $-\frac{1}{\beta}[\textbf{F}^{\prime}]_{c^{\prime}}$, does not play role in our calculations, so we can take it as a constant. From Eq. \ref{eq:45}, we know that $\textbf{Z}_{ij}$ for a quenched configuration is a function of ${\bigg \langle \sum_{k} s_{jk} s_{ki} e_{jk}e_{ki} \bigg\rangle}_{Z^{\prime}}$. We make an approximation in this step and suppose that the above mentioned quantity that appears in ${\textbf{Z}}_{ij}$ is approximated by the number of triangles, i.e. $m$ established on existing tie $\{i,j\}$, multiplied by the mean of two-stars, i.e. $\langle ss \rangle$ in that specific quenched configuration. So in Eq. \eqref{eq:14} the sum over $\{c_{ij}\}$ is approximated by the sum over all possible number of triangles ($m$) multiplied by the probability of establishing $m$ triangles on the tie $\{i,j\}$, i.e. $P(m)$. So we have:
\begin{equation}\label{eq:16}
\begin{aligned}
{[\textbf{F}]_{c}}&= -\frac{1}{\beta} \bigg ([\textbf{F}^{\prime}]_{c^{\prime}}+  \sum_{m=0}^{N-2} P(m) \ln \textbf{Z}_{ij}(m)   \bigg)\\
&= -\frac{1}{\beta}\Bigg ( [\textbf{F}^{\prime}]_{c^{\prime}} + \sum_{m=0}^{N-2} P(m)\ln \bigg( \cosh \big( \frac{\beta}{N} m {\big\langle ss \big \rangle}_{Z'} +\beta h_{ij} \big) \bigg) \Bigg).
\end{aligned}
\end{equation}
%%%%%%%%%%%%%%%%%%%%%%%%%%%%%%%%%%%%%%%%%%%%%%%%%%%%%%%%%%%%%%%%%%%%%%%%%%%%%%%%%%%%%%%%%%
\section{Configurational free energy: two-body Hamiltonian approach}
\label{appendix:c}
Like the approach we applied for calculating \textit{configurational free energy} regarding the one-body Hamiltonian partition function, for calculating \textit{configurational free energy} of the system regarding the two-body Hamiltonian partition function, we separate $\{ik, kj\}$ part and follow a similar procedure. Therefore we have: 
\begin{equation}\label{eq:17}
\begin{aligned}
{[\textbf{F}]_{c}}&=-\frac{1}{\beta}\bigg ( [\textbf{F}^{\prime \prime}]_{c^{\prime\prime}}+ \sum_{\{c_{ik, kj}\}} \ln \textbf{Z}_{ik, kj}\bigg) \\
&=-\frac{[\textbf{F}^{\prime \prime}]_{c^{\prime \prime}}}{\beta}- \frac{1}{\beta} \sum_{m_{1}=0}^{N-3} \sum_{m_{2}=0}^{N-3} P(m_{1}) P(m_{2}) \ln \big[ \textbf{Z}_{ik, kj} (m_{1}, m_{2}) \big] \\
&= -\frac{[\textbf{F}^{\prime \prime}]_{c^{\prime \prime}} }{\beta} -\frac {1}{\beta} \sum_{m_{1}=0}^{N-3} \sum_{m_{2}=0}^{N-3} P(m_{1}) P(m_{2})\ln \bigg[ \\
& e^{\frac{\beta}{N}\big(m_{1} \langle s_{il}s_{lk}\rangle_{Z^{\prime}} +m_{2} \langle s_{kl}s_{lj}\rangle_{Z^{\prime \prime}}+\langle s_{ij}\rangle_{Z^{\prime \prime}}\big)+\beta h_{ik,kj}  } \\
&+ e^{\frac{\beta}{N}\big(m_{1} \langle s_{il}s_{lk}\rangle_{Z^{\prime \prime}} -m_{2}\langle s_{kl}s_{lj}\rangle_{Z^{\prime \prime}}- \langle s_{ij}\rangle_{Z^{\prime \prime}}\big) -\beta h_{ik,kj}  }\\
&+e^{\frac{\beta}{N} \big(-m_{1} \langle s_{il}s_{lk}\rangle_{Z^{\prime \prime}} +m_{2} \langle s_{kl}s_{lj}\rangle_{Z^{\prime \prime}}- \langle s_{ij}\rangle_{Z^{\prime \prime}}\big ) -\beta h_{ik,kj} }\\
&+e^{\frac{\beta}{N}\big(-m_{1} \langle s_{il}s_{lk}\rangle_{Z^{\prime \prime}} -m_{2}\langle s_{kl}s_{lj} \rangle_{Z^{\prime \prime}}+ \langle s_{ij}\rangle_{Z^{\prime \prime}}\big) +\beta h_{ik,kj}  }\bigg]\\
&=-\frac{[\textbf{F}^{\prime \prime}]_{c^{\prime \prime}}}{\beta}-\frac{1}{\beta}\ln \bigg[ e^{\beta(2  p^2 \frac{(N-3)}{N}\langle ss\rangle +\frac{\langle s \rangle}{N} +h_{ik,kj})} + 2e^{{\beta}(\frac{-\langle s \rangle}{N}  -h_{ik,kj})} + e^{\beta(-2 p^2 \frac{(N-3)}{N}\langle ss\rangle +\frac{\langle s \rangle}{N} +h_{ik,kj})} \bigg],\\
\end{aligned}
\end{equation}
%%%%%%%%%%%%%%%%%%%%%%%%%%%%%%%%%%%%%%%%%%%%%%%%%%%%%%%%%%%%%%%%%%%%%%%%%%%%%%%%%%%%%%%%%%
Where, $m_{1}$ and $m_{2}$ are the number of triangles established on ties $\{i,k\}$ and $\{k,j\}$, respectively not including nodes $j$ and $i$.
The homogeneity of Erd\H{o}s-R\'enyi random graph lets us to assume $m_{1}= m_{2}$. From statistical mechanics we know that the mean of two-stars, $\langle s_{ik} s_{kj}\rangle$, is the first derivative of free energy with respect to the field $h_{ik,kj}$:
\begin{equation}\label{eq:18}
\begin{aligned}
\langle s_{ik} s_{kj}\rangle&=-\frac{\partial [\textbf{F}]}{\partial h_{ik, kj}} {\Bigg|}_{h_{ik,kj}=0}\\
&=-\frac{\partial( -\frac{[\textbf{F}^{\prime\prime}]_{c^{\prime\prime}}}{\beta})}{\partial h_{ik,kj}}{\Bigg|}_{h_{ik,kj}=0}+
\frac{1}{\beta} \frac{\partial}{\partial h_{ik,kj}} \ln \bigg[ e^{\beta(2  p^2 \frac{(N-3)}{N}\langle ss\rangle +\frac{\langle s \rangle}{N} +h_{ik,kj})} + 2e^{{\beta}(\frac{-\langle s \rangle}{N}  -h_{ik,kj})} + e^{\beta(-2 p^2 \frac{(N-3)}{N}\langle ss\rangle +\frac{\langle s \rangle}{N} +h_{ik,kj})} \bigg] {\Bigg|}_{h_{ik,kj}=0}\\
&=0+ \frac{\bigg( e^{\beta(2p^2 \frac{N-3}{N}\langle ss\rangle )} - 2e^{{\beta}(-2\frac{\langle s \rangle}{N} )} + e^{\beta(-2 p^2 \frac{N-3}{N} \langle ss\rangle  )} \bigg)}{\bigg( e^{\beta(2 \frac{N-3}{N} p^2 \langle ss\rangle )} + 2e^{{\beta}(-2\frac{\langle s \rangle}{N} )} + e^{\beta(-2 p^2 \frac{N-3}{N}\langle ss\rangle  )} \bigg)}.
\end{aligned}
\end{equation}
%%%%%%%%%%%%%%%%%%%%%%%%%%%%%%%%%%%%%%%%%%%%%%%%%%%%%%%%%%%%%%%%%%%%%%%%%%%%%%%%%%%%%%%%%%
\section{Graphical representation for self-consistent equation \ref{eq:13} for different connection probability}
\label{appendix:a}
In Fig. \ref{fig:4} we have illustrated the graphical representation of Eq. \eqref{eq:13} for different connection probabilities. There is a single stable fixed point for $T<T_{cold}$ and $T>T_{hot}$ which represent the balanced and random phase respectively. When $T_{cold}<T<T_{hot}$ we have two stable fixed points that represent the coexistence regions. Therefore the number of intersections of Eq. \ref{eq:13} truly demonstrates the phase of our random network. 
\begin{figure}[htbp] % "[htbp]" placement specifier just for this example
	\begin{subfigure}{0.3\textwidth}
		\includegraphics[width=\linewidth]{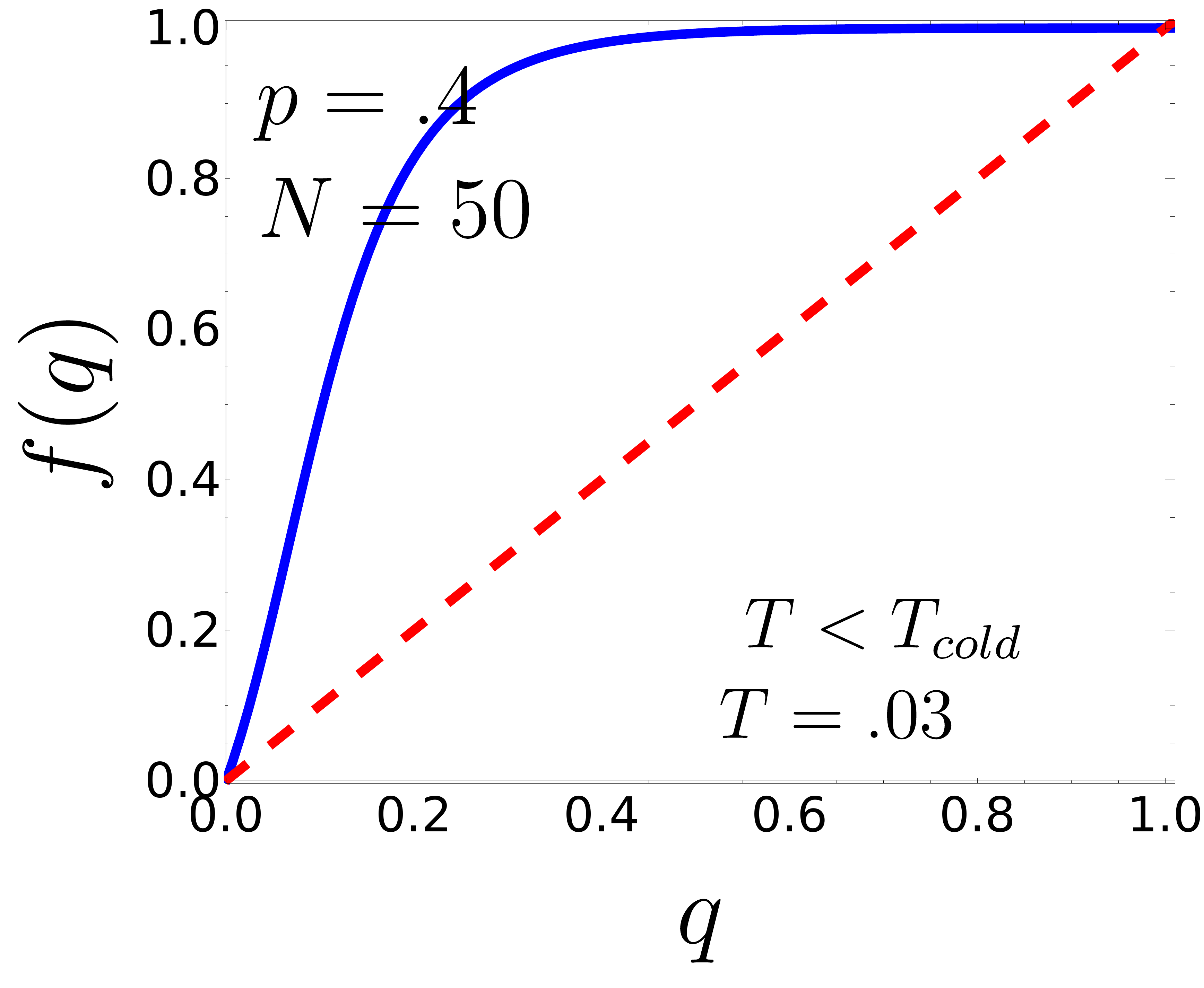}
		%\caption{} \label{fig:a}
	\end{subfigure}
	\begin{subfigure}{0.3\textwidth}
		\includegraphics[width=\linewidth]{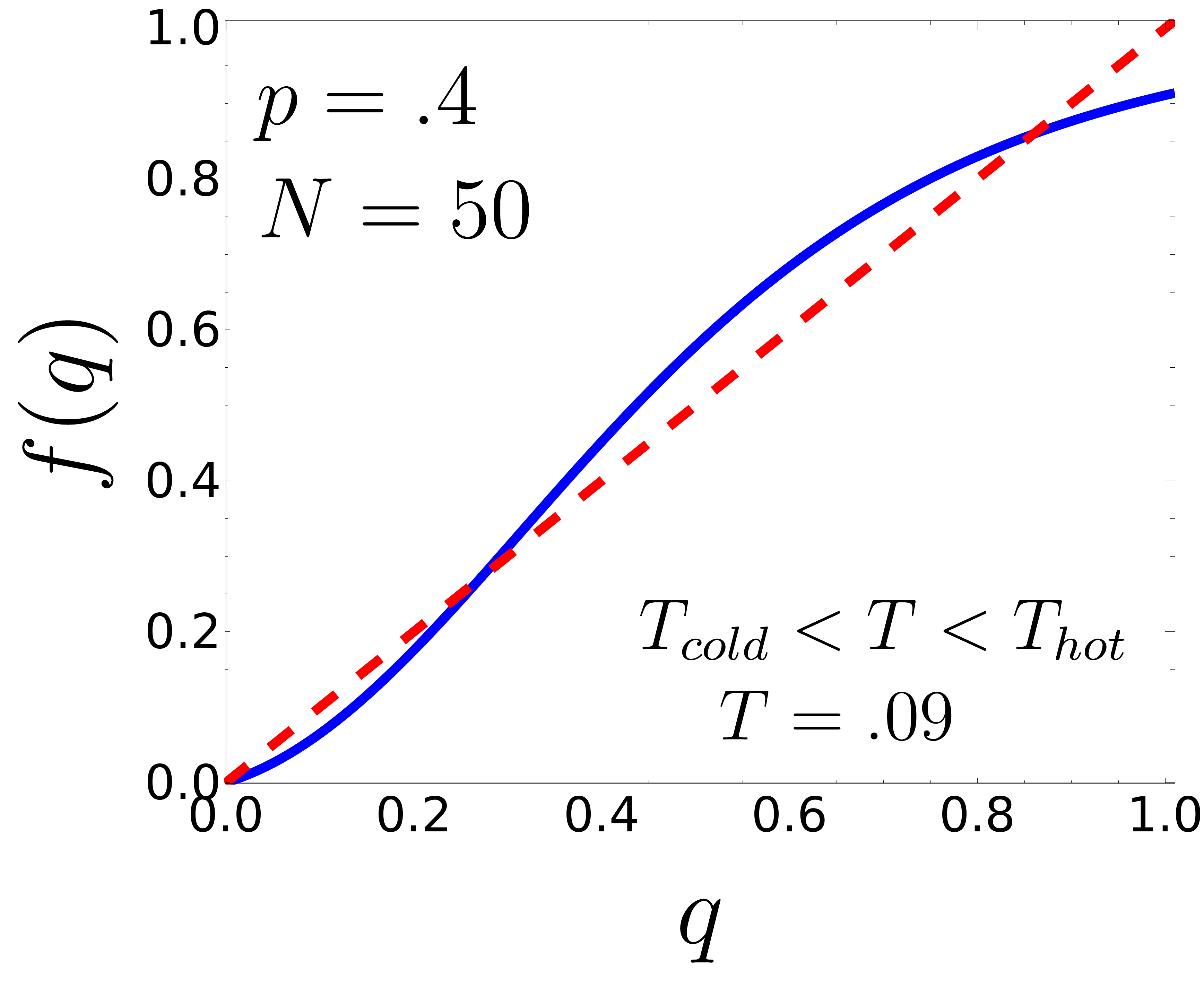}
		%\caption{} \label{fig:b}
	\end{subfigure}
	\begin{subfigure}{0.32\textwidth}
		\includegraphics[width=\linewidth]{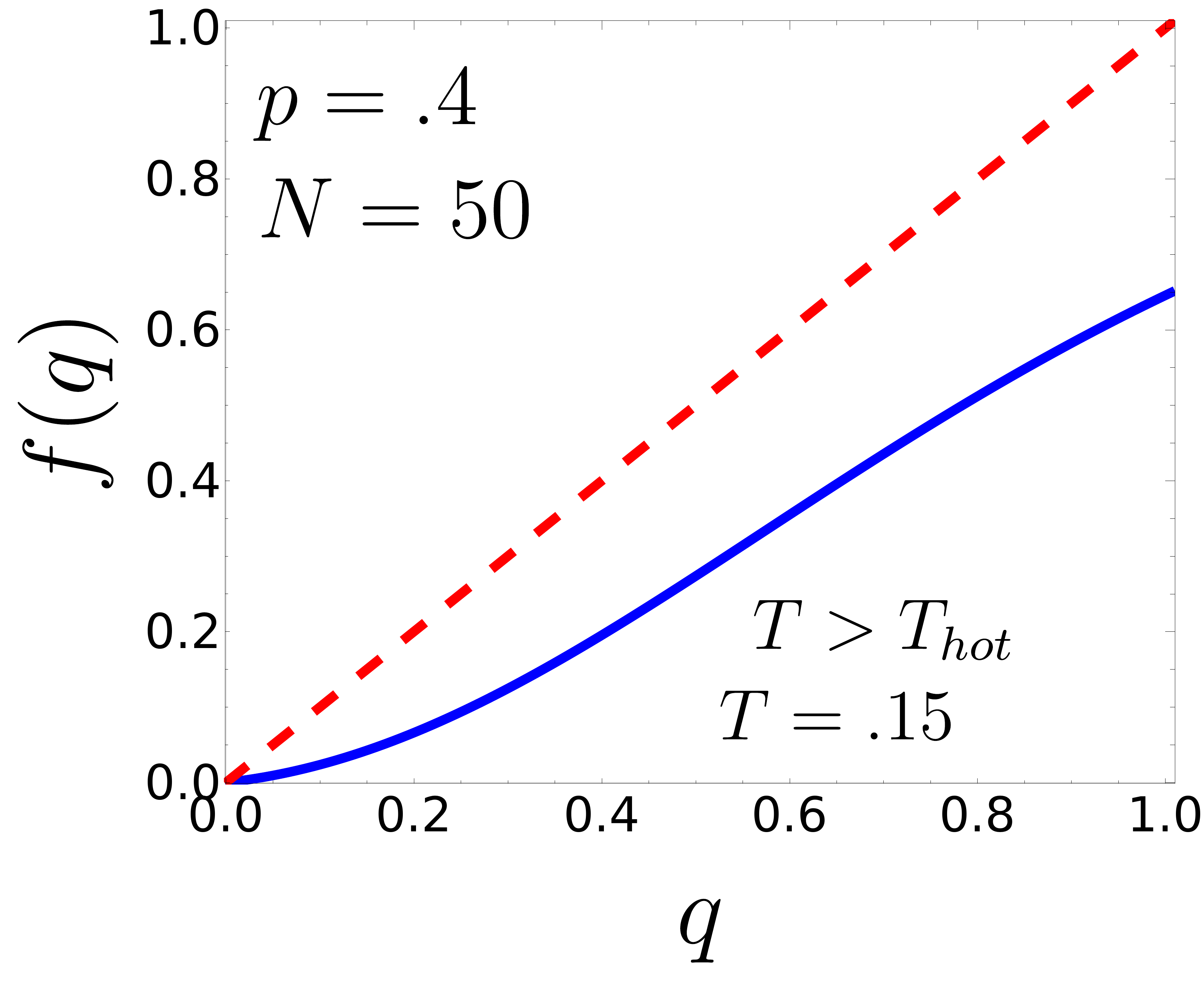}
		%\caption{} \label{fig:c}
	\end{subfigure}
	\medskip
	\begin{subfigure}{0.3\textwidth}
		\includegraphics[width=\linewidth]{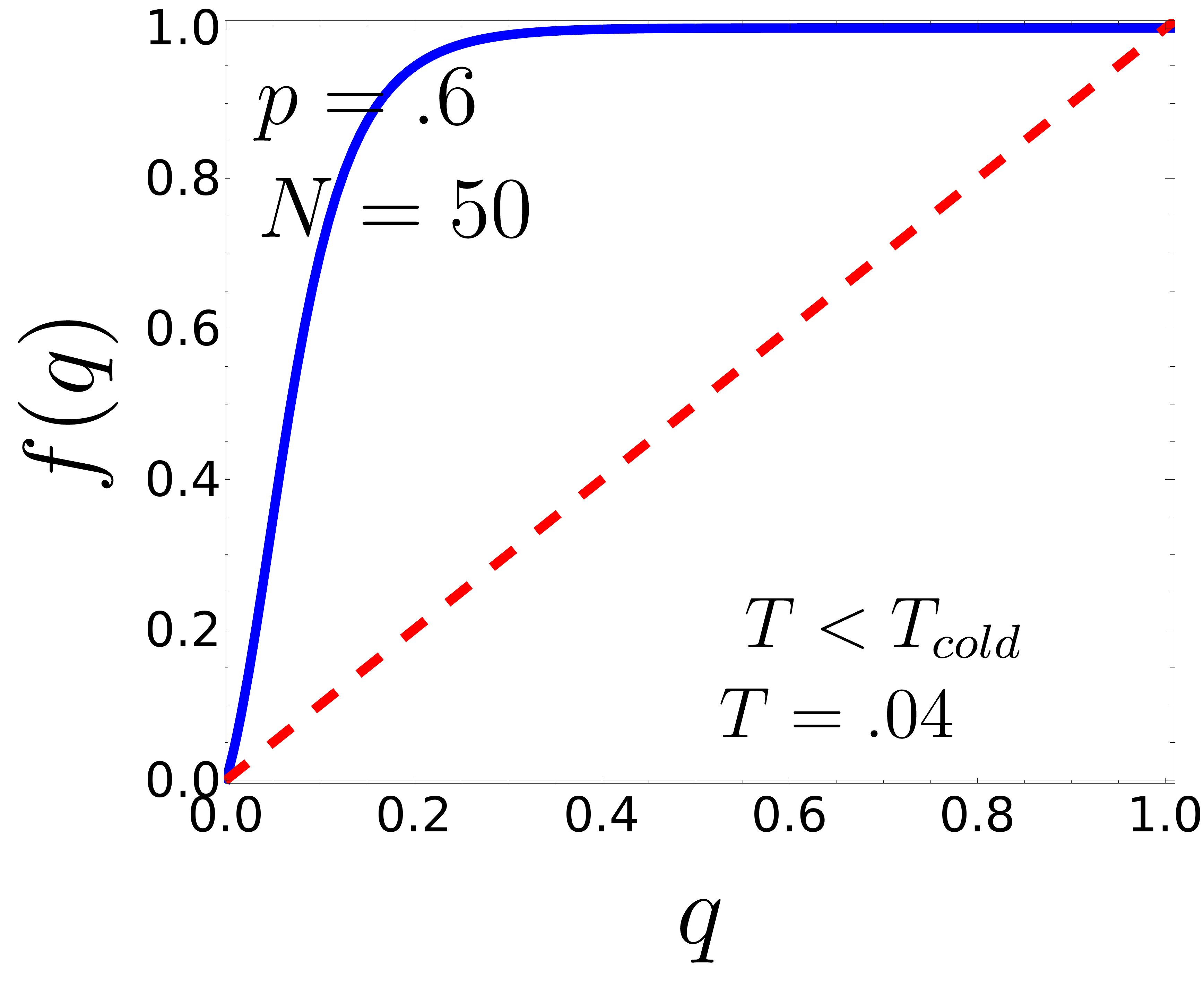}
		%\caption{} \label{fig:d}
	\end{subfigure}
	\begin{subfigure}{0.3\textwidth}
		\includegraphics[width=\linewidth]{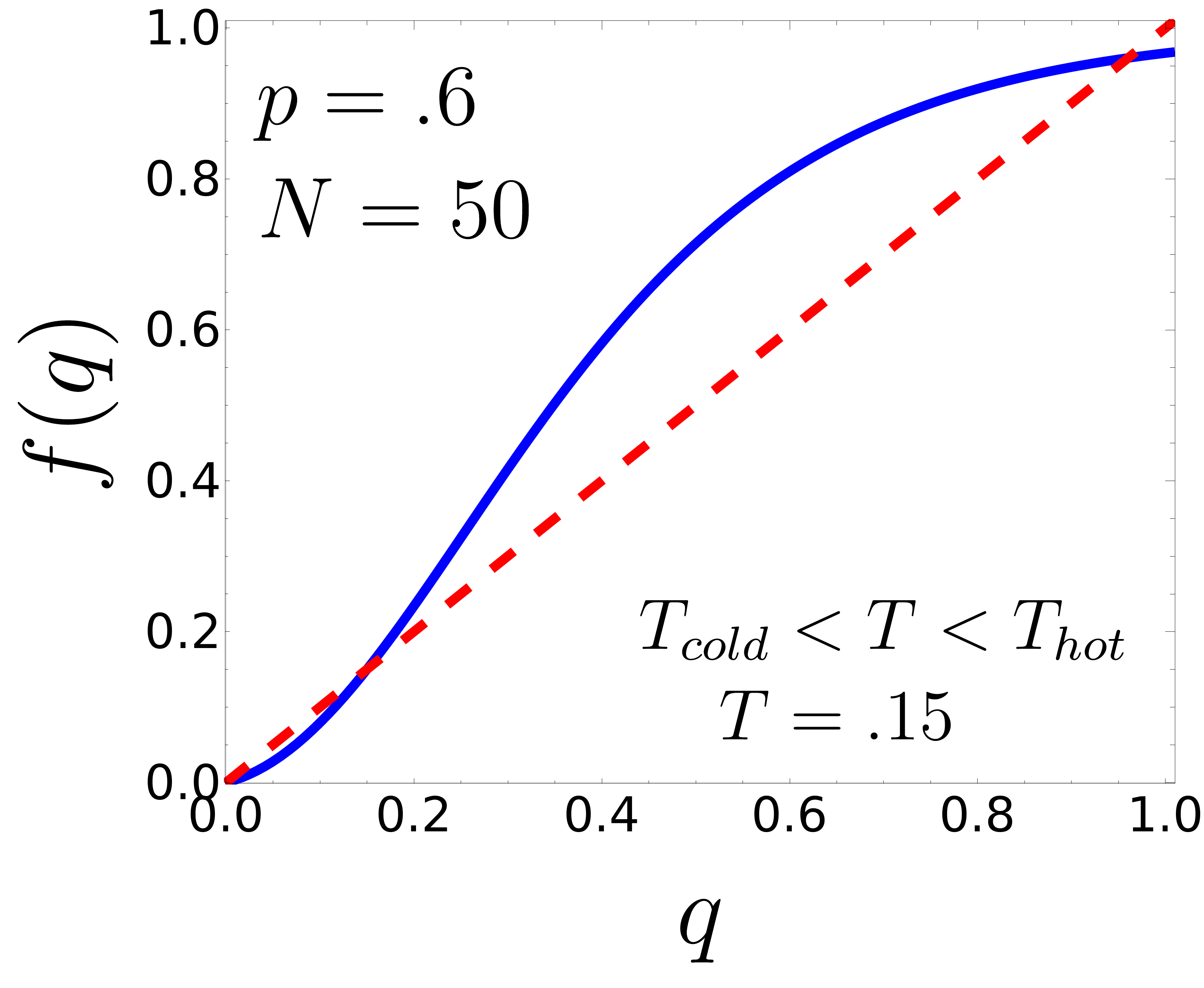}
	\end{subfigure}
	\begin{subfigure}{0.3\textwidth}
		\includegraphics[width=\linewidth]{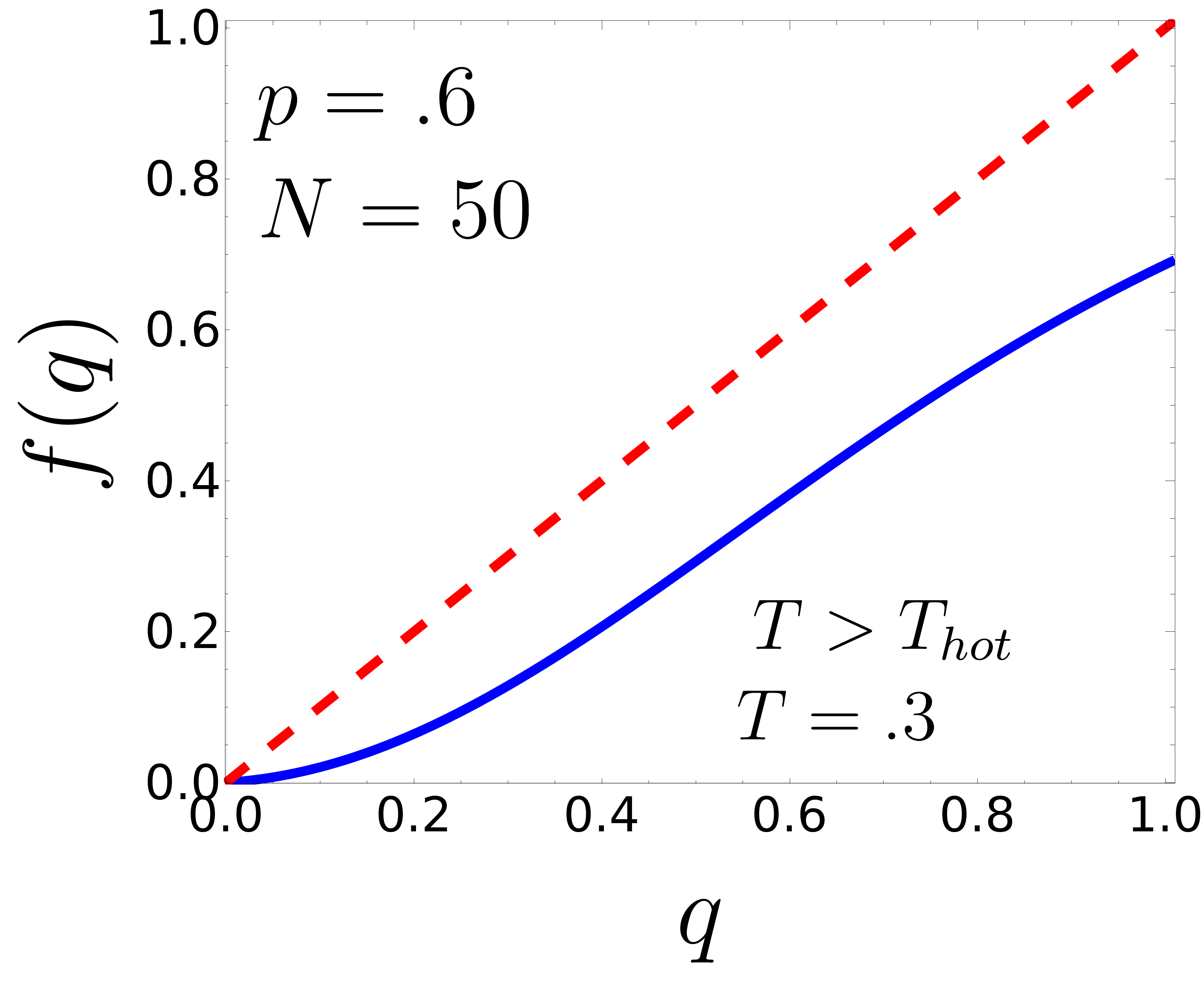}
		%\caption{} \label{fig:f}
		
	\end{subfigure}
	\begin{subfigure}{0.3\textwidth}
		\includegraphics[width=\linewidth]{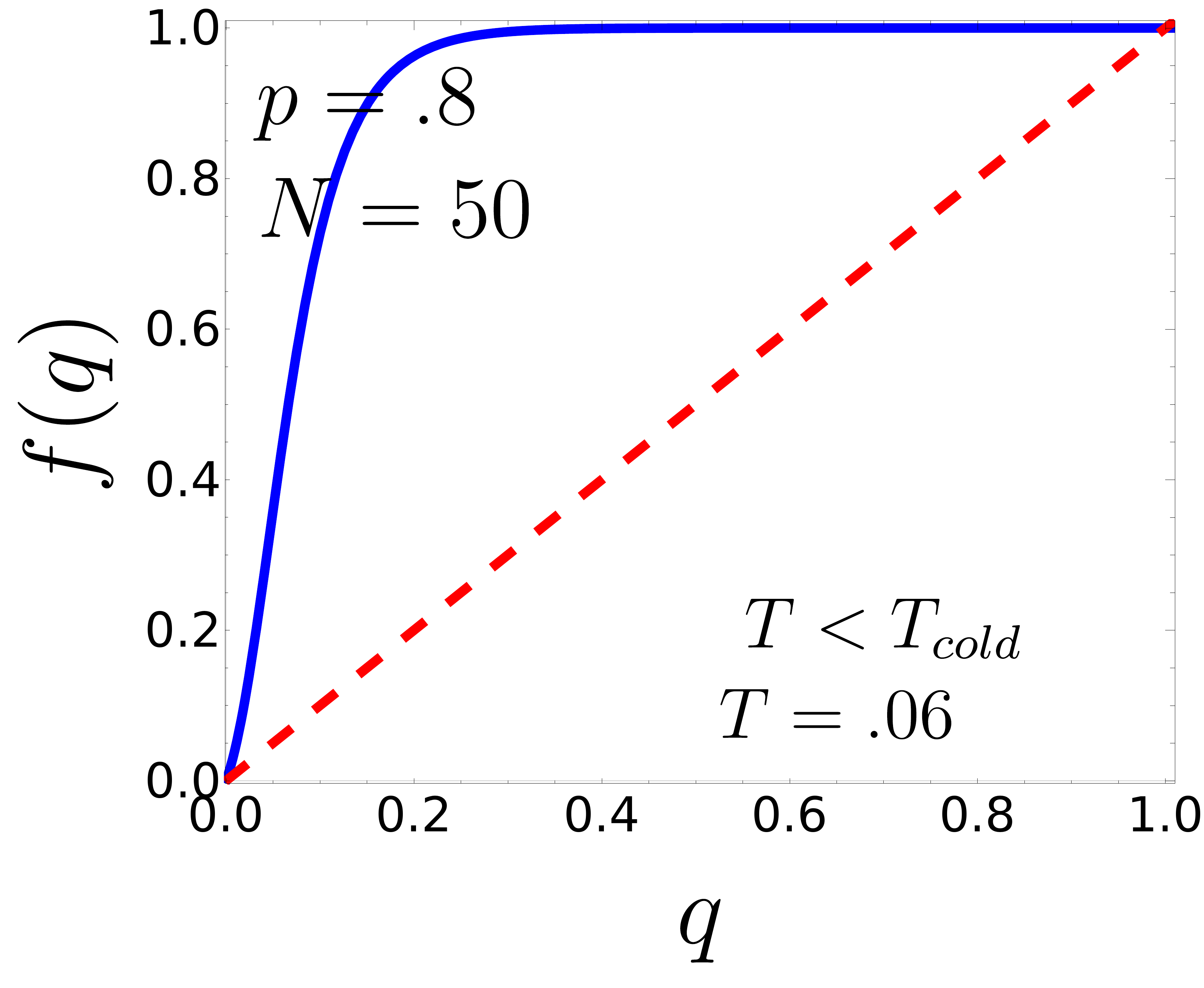}
		%\caption{} \label{fig:f}
		
	\end{subfigure}
	\medskip
	\begin{subfigure}{0.3\textwidth}
		\includegraphics[width=\linewidth]{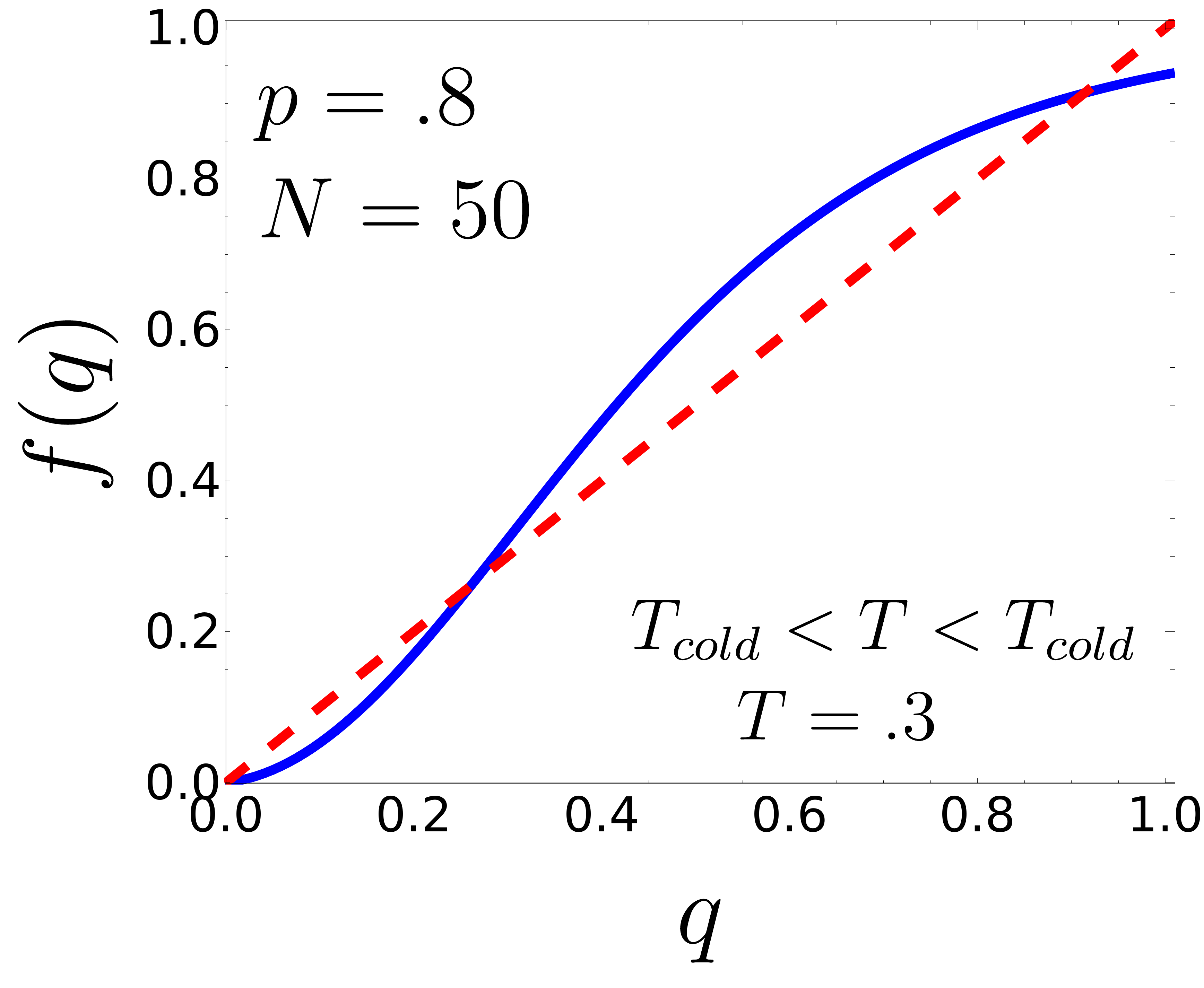}
		%	\caption{} \label{fig:f}
		
	\end{subfigure}
	\begin{subfigure}{0.3\textwidth}
		\includegraphics[width=\linewidth]{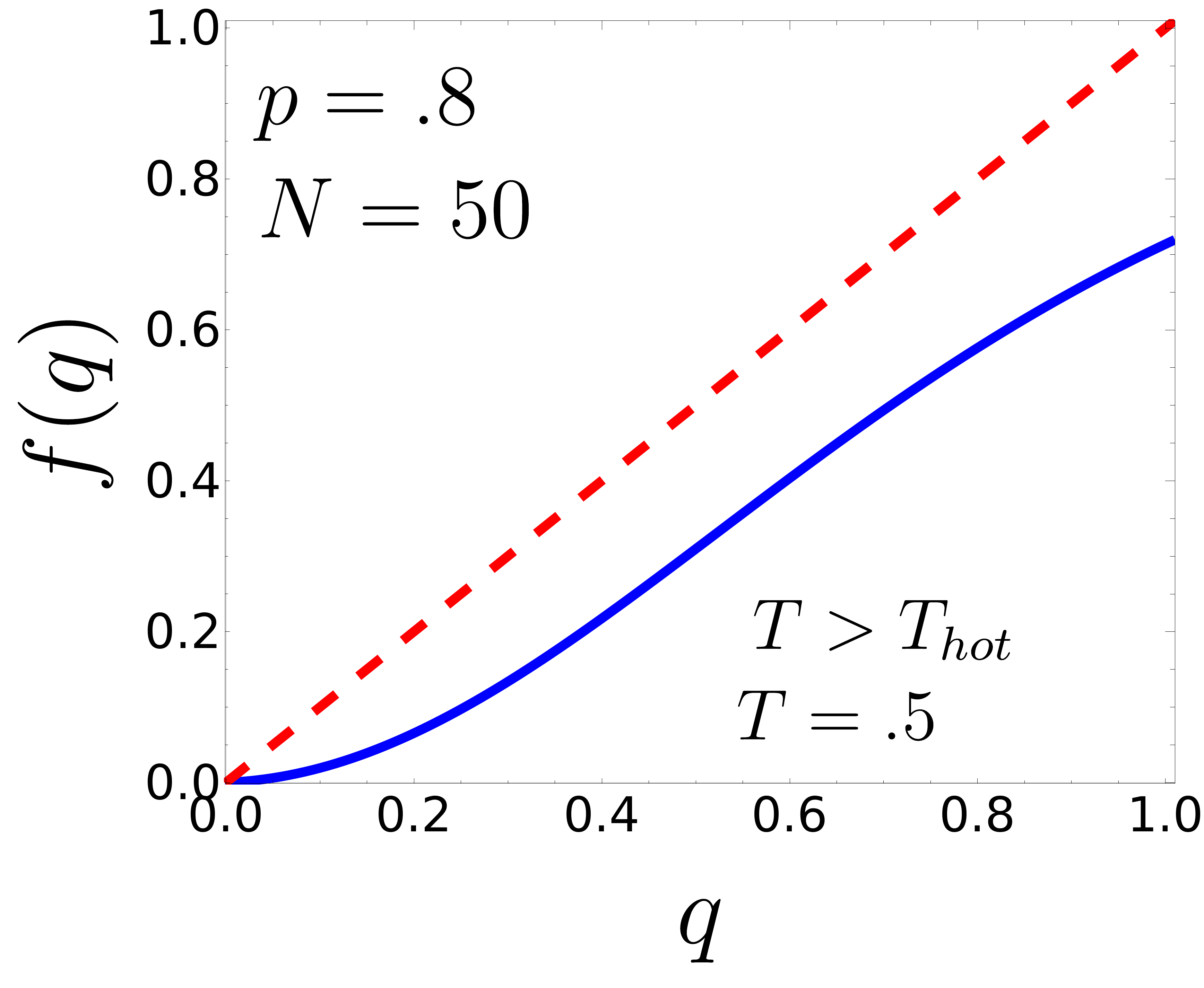}
		%\caption{} \label{fig:f}
		
	\end{subfigure}
	\begin{subfigure}{0.3\textwidth}
		\includegraphics[width=\linewidth]{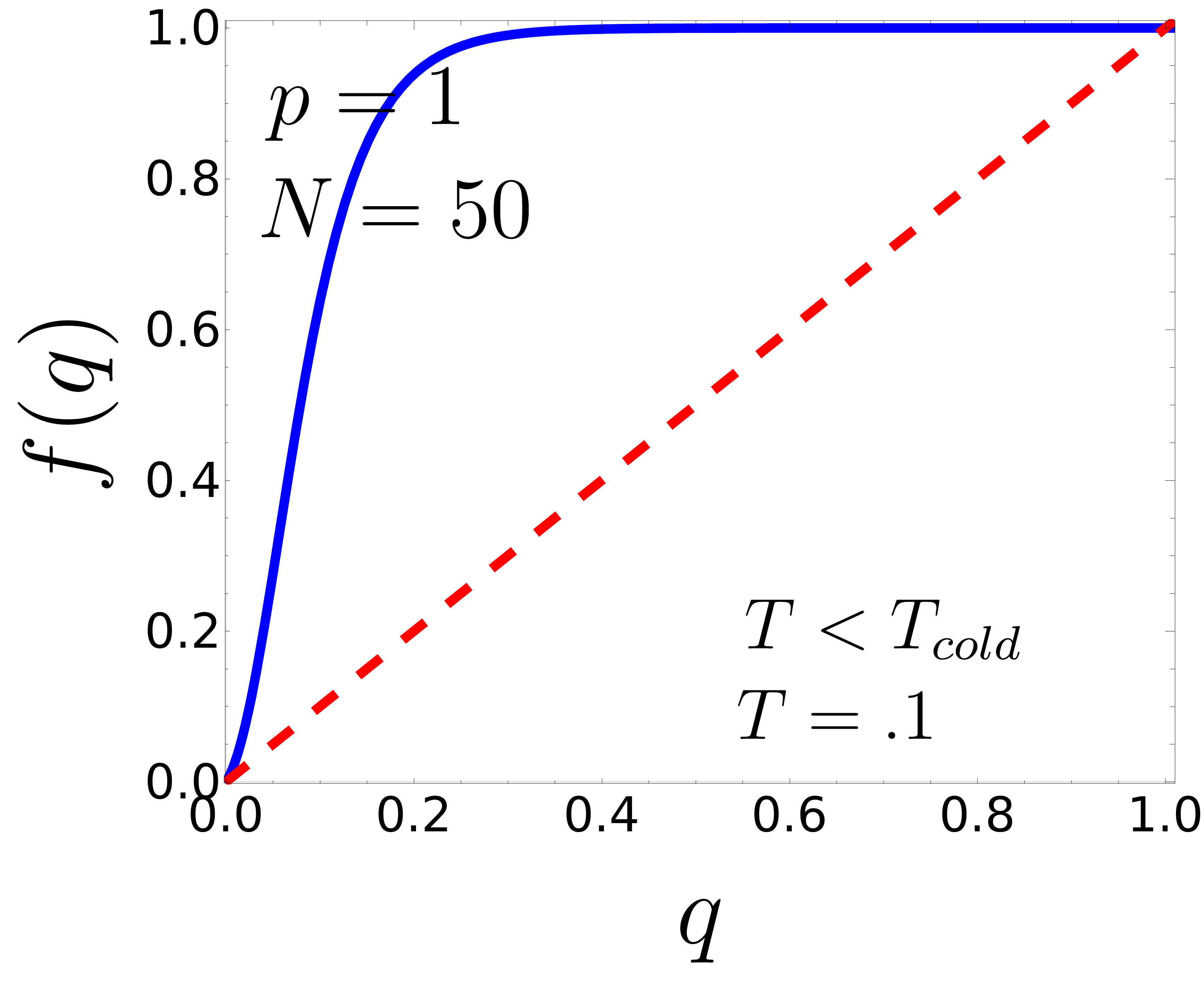}
		%\caption{} \label{fig:f}
		
	\end{subfigure}
	\medskip
	\begin{subfigure}{0.3\textwidth}
		\includegraphics[width=\linewidth]{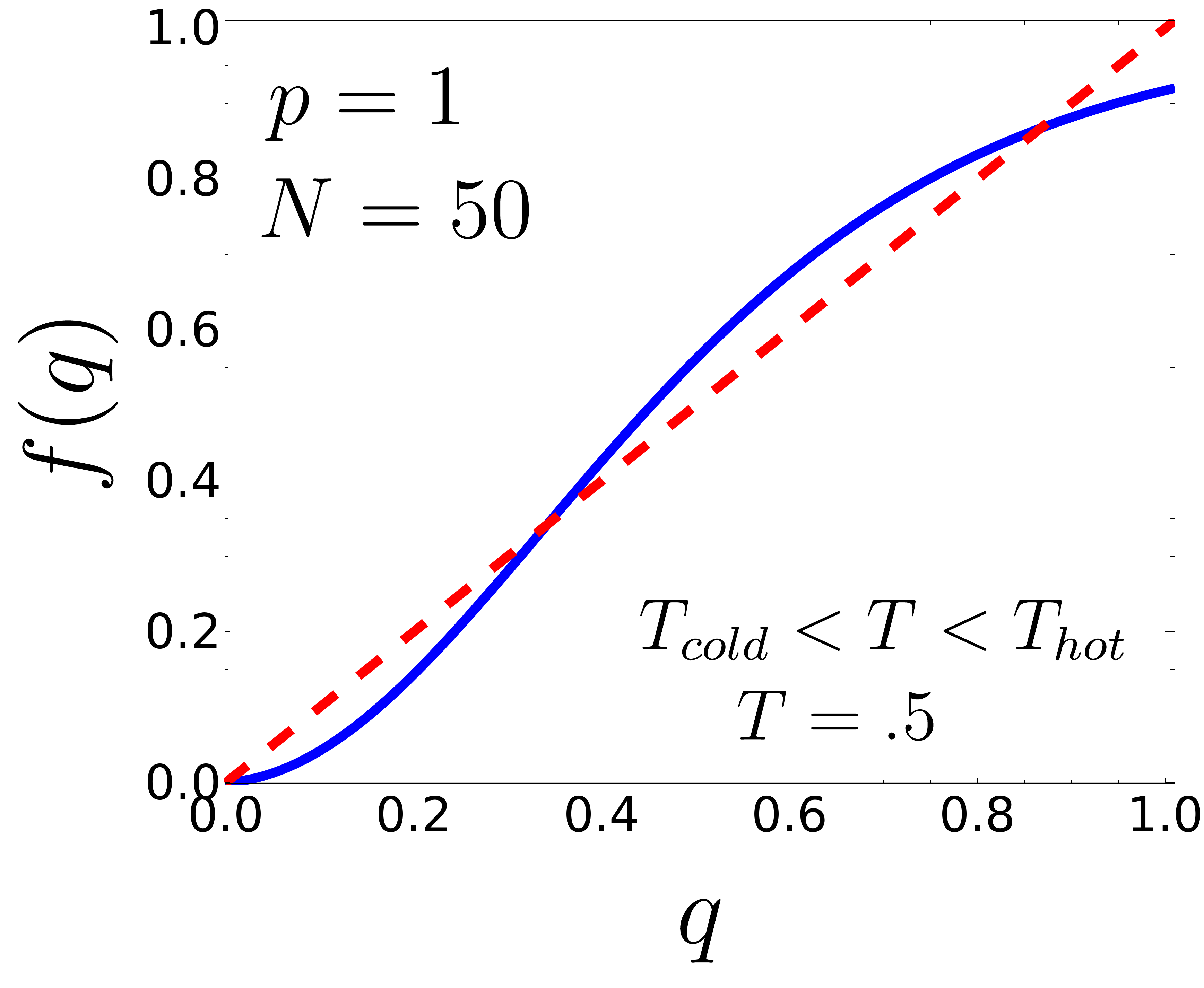}
		%	\caption{} \label{fig:f}
		
	\end{subfigure}
	\begin{subfigure}{0.3\textwidth}
		\includegraphics[width=\linewidth]{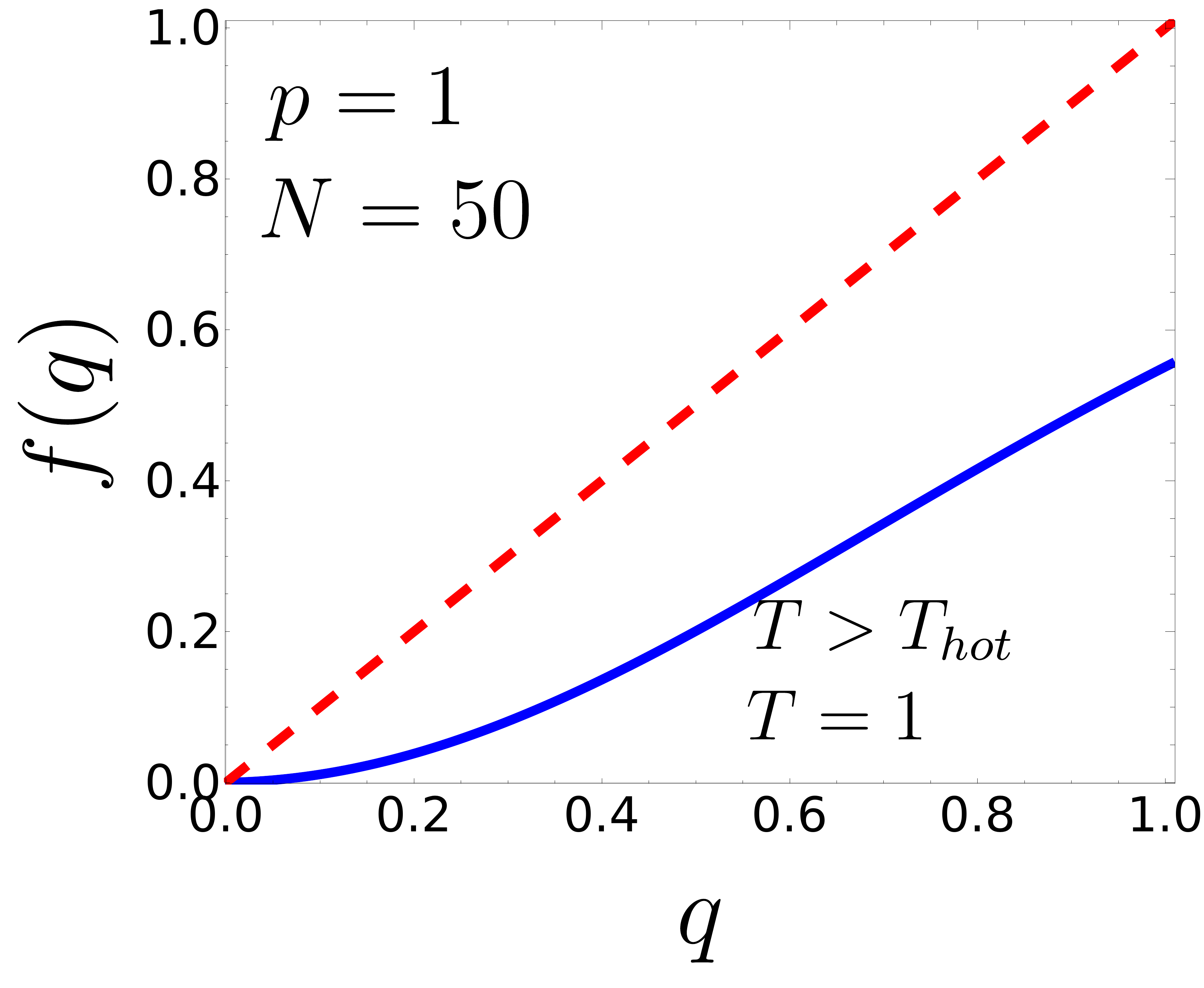}
		%\caption{} \label{fig:f}
	\end{subfigure}
	\caption{Graphical representation of self-consistent Eq. \ref{eq:13} for four connection probabilities $p=1$, $p=.8$ , $p=.6$ and $p=.4$ in three distinct phases for an  Erd\H{o}s-R\'enyi graph of size $N=50$. } \label{fig:4}
\end{figure}
%%%%%%%%%%%%%%%%%%%%%%%%%%%%%%%%%%%%%%%%%%%%%%%%%%%%%%%%%%%%%%%%%%%%%%%%%%%%%%%%%%%%%%
\section{Calculating Cold Critical Temperature}
	\label{appendix:d}
	As we discussed earlier, $T_{cold}$ is where the coexistence region starts appearing. We know that in coexistence region $q^{*}=0$ is a stable fix-pint. Hence, for calculating $T_{cold}$, we need to drive the first derivative of the right hand of self-consistent equation $q=f(q)$ in $q^{*}=0$. We need to have it as a stable fixed point so we should have $f^{\prime} (q^{*}=0)<1$. So we have:
	\begin{eqnarray}\label{eq:19}
	\begin{aligned}
	\frac{\partial f(q)}{\partial q} {\bigg|}_{q=0}&= \frac{\bigg( 4\beta p^{2}(\frac{N-3}{N}) \sinh [2\beta p^{2} (\frac{N-3}{N}) q]+ 4{\beta}^{2} p^{2}(\frac{N-2}{N^2}) (1-{\tanh}^{2} [\beta p^{2}(\frac{N-2}{N}) q]) e^{-2 \beta \frac{\langle s \rangle}{N} } \bigg) \bigg( 2\cosh[2 (\frac{N-3}{N}) \beta p^{2}q] +2 e^{-2\beta \langle s \rangle}\bigg)}
	{\bigg( 2\cosh [2 \beta p^{2}\frac{N-3}{N} q] +2e^{-2\beta \frac{\langle s \rangle}{N} } \bigg)^{2}} {\bigg|}_{q=0}\\
	&\frac{\bigg( 4\beta p^{2}(\frac{N-3}{N}) \sinh [2\beta p^{2} (\frac{N-3}{N}) q]- 4{\beta}^{2} p^{2} (\frac{N-2}{N^2}) (1-{\tanh}^{2} [\beta p^{2}(\frac{N-2}{N}) q]) e^{-2 \beta \frac{\langle s \rangle}{N} } \bigg) \bigg( 2\cosh[2 (\frac{N-3}{N}) \beta p^{2}q] -2 e^{-2\beta \langle s \rangle}\bigg)}
	{\bigg( 2\cosh [2 \beta p^{2}\frac{N-3}{N} q] +2e^{-2\beta \frac{\langle s \rangle}{N} } \bigg)^{2}} {\bigg|}_{q=0}\\
	&={\beta}^{2} {p}^{2} (\frac{N-2}{N^2})
	\end{aligned}
	\end{eqnarray}
	Now if we apply $f^{\prime} (q^{*}=0)<1$ we have:
	\begin{equation*}
	T_{cold}=p \sqrt{\frac{N-2}{N^{2}}}
	\end{equation*}
%%%%%%%%%%%%%%%%%%%%%%%%%%%%%%%%%%%%%%%%%%%%%%%%%%%%%%%%%%%%%%%%%%%%%%%%%%%%%%%%%%%%%
\end{appendices}
\end{widetext}

\end{document}